\documentclass[reprint,aps,prx,10pt,floatfix]{revtex4-2}
\usepackage{amsmath}
\usepackage{physics}
\usepackage{bm}
\usepackage{upgreek}
\usepackage{graphicx}
\usepackage[T1]{fontenc}
\usepackage{multirow}
\usepackage[colorlinks=true, allcolors=blue]{hyperref}

\begin{document}

\title{
Autonomously Designed Pulses for Precise, Site-Selective Control of Atomic Qubits
}
\author{Sanghyo Park}
\affiliation{Department of Physics, Korea Advanced Institute of Science and Technology (KAIST), Daejeon 34141, South Korea}
\author{Seuk Lee}
\affiliation{Department of Physics, Korea Advanced Institute of Science and Technology (KAIST), Daejeon 34141, South Korea}
\author{Keunyoung Lee}
\affiliation{Department of Physics, Korea Advanced Institute of Science and Technology (KAIST), Daejeon 34141, South Korea}
\author{Minhyeok Kim}
\affiliation{Department of Physics, Korea Advanced Institute of Science and Technology (KAIST), Daejeon 34141, South Korea}
\author{Donggyu Kim}
\thanks{Corresponding author (e-mail: dngkim@kaist.ac.kr)}
\affiliation{Department of Physics, Korea Advanced Institute of Science and Technology (KAIST), Daejeon 34141, South Korea}
% \date{\today}

\begin{abstract}
Quantum computers based on cold-atom arrays offer long-lived qubits with programmable connectivity, yet their progress toward fault-tolerant operation is limited by the relatively low fidelity of site-selective local control. We introduce an artificial-intelligence (AI) framework that overcomes this limitation. Trained on atom-laser dynamics, a deep neural network autonomously designs composite pulses that improve local control fidelities tenfold while remaining compatible with existing control hardware. We further demonstrate the robustness of these pulses against optical aberrations and beam misalignment. This approach establishes AI-trained pulse compilation for high-fidelity qubit control and can be readily extended to other atom-like platforms, such as trapped ions and solid-state color centers.
\end{abstract}

\maketitle

\section{Introduction}

Neutral-atom arrays are emerging as a leading platform for large-scale quantum information processing~\cite{intro_processor_scholl_quantum_2021, intro_processor_shaw_benchmarking_2024, intro_processor_ebadi_quantum_2022, intro_processor_bluvstein_logical_2024, intro_processor_manovitz_quantum_2025}. Each atom trapped in an optical tweezer hosts long-lived qubit states that are precisely manipulated using near-resonant laser illuminations~\cite{intro_qubit_norcia_seconds-scale_2019, intro_processor_barnes_assembly_2022, intro_qubit_park_technologies_2024}. Control fidelity approaching 0.9999 for single-qubit gates~\cite{intro_fidel_levine_dispersive_2022} and 0.999 for two-qubit gates~\cite{intro_fidel_evered_high-fidelity_2023} has been demonstrated with global laser illumination across entire atom arrays. However, extending this level of precision to site-selective, individual qubit control remains a major challenge on the path to fully programmable, fault-tolerant quantum computing~\cite{intro_processor_bluvstein_logical_2024}. The global illuminations combined with the atom shuttling~\cite{intro_barredo_atom-by-atom_2016} enable individual qubit control~\cite{intro_global_bluvstein_quantum_2022, intro_qubit_shaw_multi-ensemble_2024}, but this approach limits the gate speed and increases the complexity of compiling quantum circuits~\cite{intro_compling_tan_compiling_2024}.

\begin{figure}
  \centering
  \includegraphics[width=0.475\textwidth]{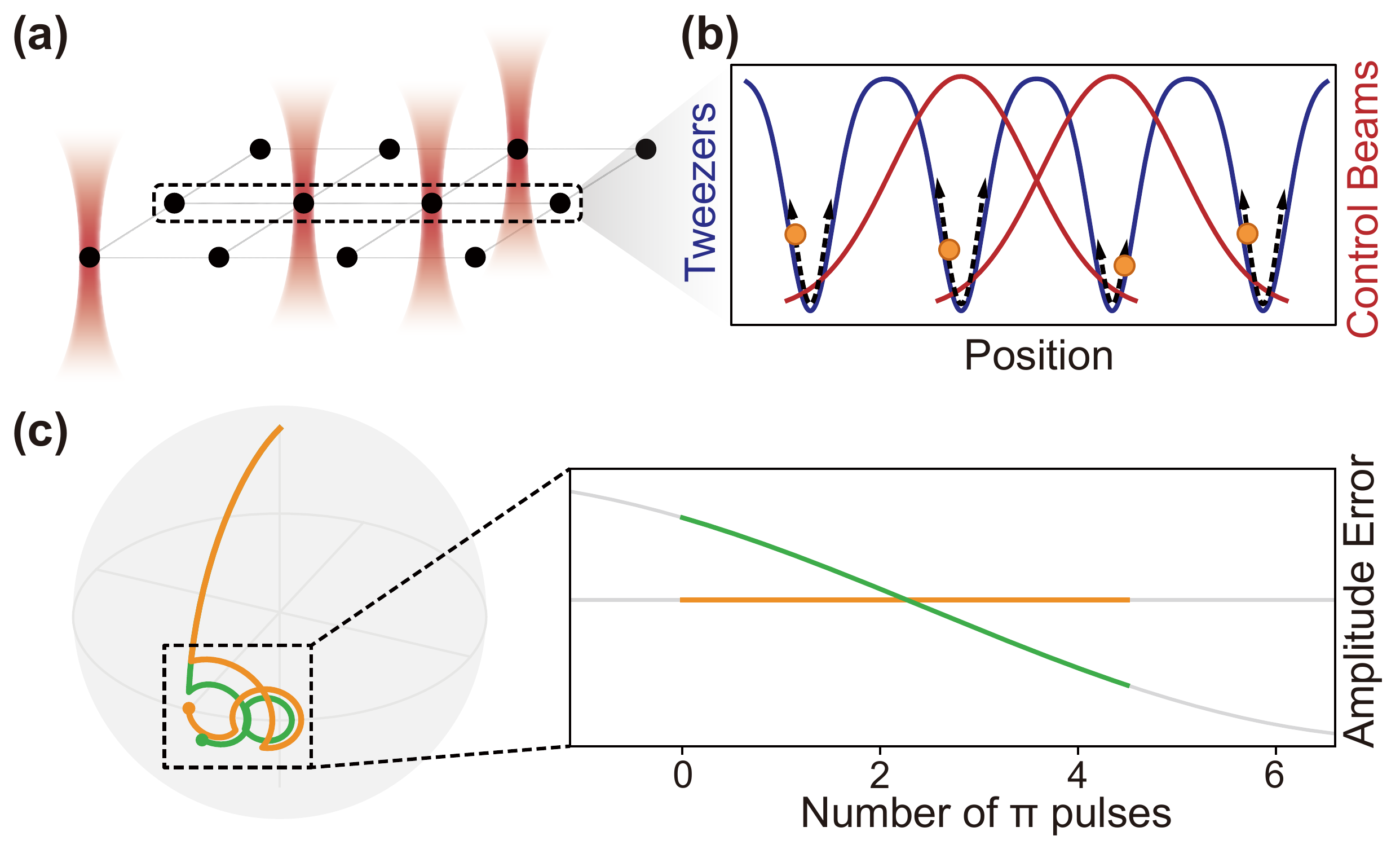}
  \caption{
    \textbf{Site-Selective Control of Atomic Qubits in Optical Tweezers}
    \textbf{(a)} Addressing target atoms with tightly focused beams enables efficient site-selective qubit control.
    \textbf{(b)} Atom motions in optical tweezers induce amplitude fluctuations in the local control field.
    \textbf{(c)} With time-varying amplitude (green), conventional composite pulses become ineffective at mitigating amplitude errors (orange).
    }
  \label{fig:intro}
\end{figure}

Using localized laser beams for individual qubit control (Fig.~\ref{fig:intro}(a)) offers several advantages, such as efficient usage of laser power, reduced off-resonant scattering~\cite{intro_fidel_levine_dispersive_2022}, lower idling errors~\cite{intro_global_bluvstein_quantum_2022}, and native circuit implementation~\cite{intro_advantage_radnaev_universal_2025}. However, this local qubit control generally shows a lower control fidelity~\cite{intro_local_lis_midcircuit_2023, intro_local_rodriguez_experimental_2025, intro_local_muniz_high-fidelity_2025, intro_processor_bluvstein_logical_2024} compared with global illumination: the atom motions within the tweezers introduce large fluctuations in the atom-laser interaction (Fig.~\ref{fig:intro}(b)), leading to the qubit control errors and decoherence~\cite{intro_motion_thompson_coherence_2013}. Increasing the beam diameter can reduce these effects, but this approach compromises the laser power efficiency and isolation from the adjacent atoms. Although motional ground-state cooling~\cite{intro_ground_spence_preparation_2022} can minimize such errors, residual motions still impose a limit on control precision.

The atom motions need not permanently restrict control precision; instead, engineered laser pulses can coherently eliminate the motion-induced effects. Techniques such as dynamical decoupling~\cite{intro_DD_khodjasteh_dynamically_2009, intro_DD_souza_robust_2012,intro_DD_harris_pulse_2016} and robust composite pulses (CPs)~\cite{intro_CP_wimperis_broadband_1994, intro_CP_brown_arbitrarily_2004, intro_CP_kabytayev_robustness_2014} effectively counteract the quasi-static errors. However, these methods become ineffective when using localized laser beams, since the atom motions induce rapidly-varying amplitude errors on the qubit control (Fig.~\ref{fig:intro}(c)). While stochastic optimizations have been explored to identify the robust pulses for specific target rotations~\cite{intro_optim_wu_composite_2023, intro_optim_xie_9992-fidelity_2023}, generalizing these approaches to arbitrary qubit rotations remains a significant challenge.

To address this challenge, we introduce machine-learned CPs for precise local control. A deep neural network (DNN), trained on a Hamiltonian describing motion-dependent atom-laser interactions, autonomously generates CPs that cancel motion-induced amplitude errors. Compared with the robust CPs currently in use, the learned sequences improve single-site gate fidelity by an order of magnitude and broaden the accessible range of qubit rotation. We further benchmark their resilience to optical aberrations and alignment errors using a programmable focus array~\cite{intro_focus_kim_large-scale_2019}. 

\section{Autonomous pulse design}

\begin{figure*}
  \centering
  \includegraphics[width=1.0\textwidth]{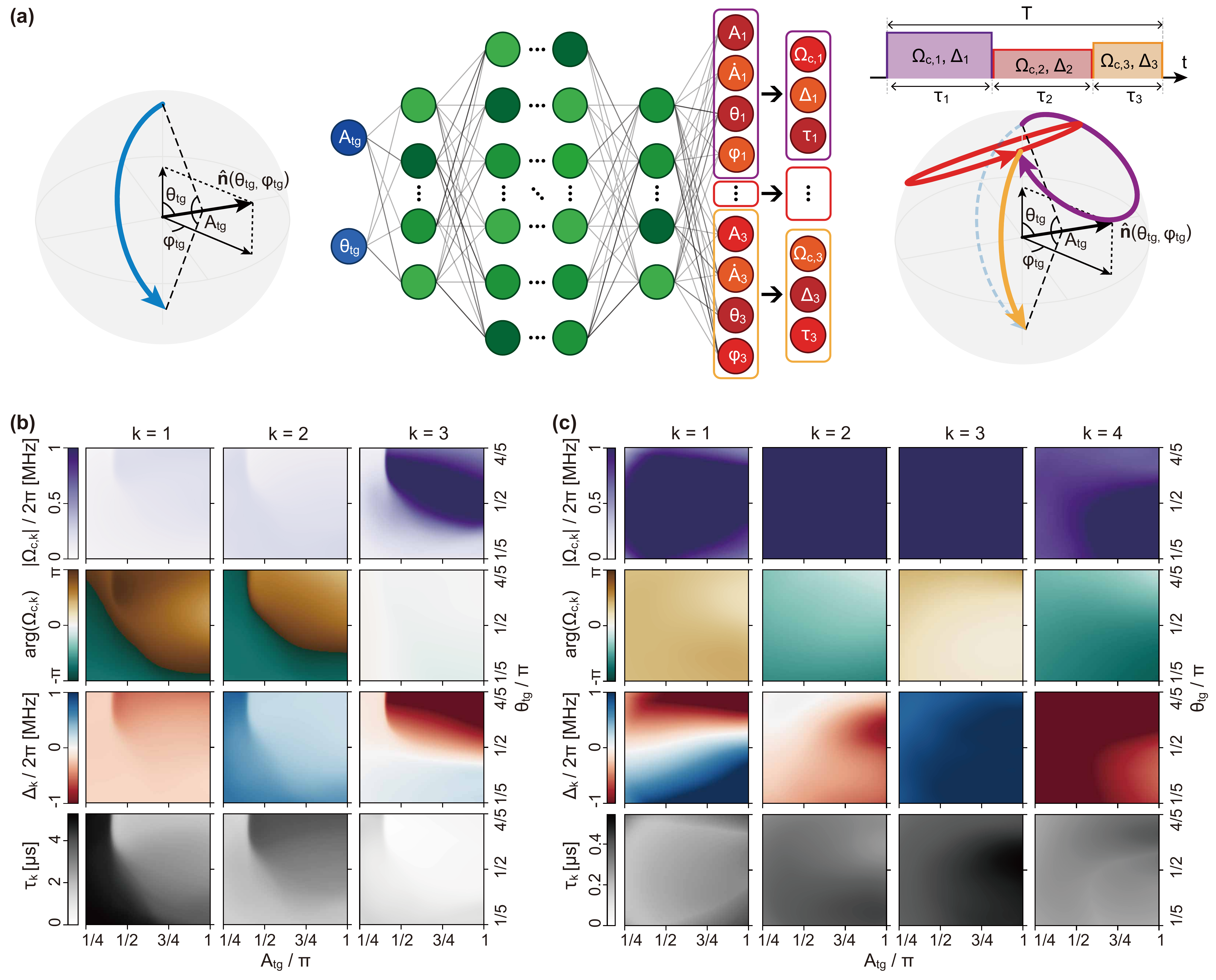}
  \caption{
    \textbf{Trained Composite Pulses}    
        \textbf{(a)} A target qubit rotation (left) on the Bloch sphere is encoded into a series of pulses (right). The trained deep neural network (middle) identifies the required pulse parameters. 
        \textbf{(b)} and \textbf{(c)} Pulse parameters of trained \textit{CP}(3) and \textit{CP}(4) implementing the target rotation ($A_\text{tg}$, $\theta_\text{tg}$), respectively.
        }
  \label{fig:trained-CPs}
\end{figure*}

Figure \ref{fig:trained-CPs}(a) shows the schematic of the DNN to produce the trained CPs. The CP comprises a series of $n$ rectangular pulses to implement target qubit control, while mitigating the motion-induced amplitude error $\epsilon(t)$. We denote the trained CP with $n$ pulses as \textit{CP}($n$); the $k$-th pulse ($k=1,...,n$) is characterized by the Rabi frequency $\Omega_{\textrm{c},k}$, detuning $\Delta_k$, and the pulse duration $\tau_k$. For each target rotation 
\begin{equation*}
    \mathcal{U} = \exp(-i\frac{A_{\rm tg}}{2}\hat{\mathbf{n}}(\theta_{\rm tg},\phi_{\rm tg})\cdot\boldsymbol{\sigma}),
\end{equation*} 
the DNN receives the desired pulse area $A_{\rm tg}$ and polar angle $\theta_{\rm tg}$ of $\hat{\mathbf{n}}$, and returns the required parameters for the $n$ constituent pulses. Adding the global phase $\phi_{\rm tg}$ to all $\Omega_{\textrm{c},k}$ completes the implementation of $\mathcal{U}$ by \textit{CP}($n$). 

The DNN is trained with reinforcement learning~\cite{DRL_niu_universal_2019, supp_fosel_reinforcement_2018}. For the reward of the learning, we use the control fidelity
\begin{equation*}
    F_{\textit{CP}(n)}(\epsilon(t)) = \frac{1}{4}\langle|{\rm Tr}(\mathcal{U}^\dagger U_{\textit{CP}(n)}(T;\epsilon(t))|^2\rangle
    \label{eq:control-fidelity}
\end{equation*}
of implementing $\mathcal{U}$ (Supp. E). Here, $U_{\textit{CP}(n)}(T;\epsilon(t))$ is the time-evolution operator due to the \textit{CP}($n$) under $\epsilon(t)$ ($T=\sum_{k=1}^n\tau_k$). $\langle ... \rangle$ refers to the ensemble average over the thermal atoms in a tweezer. The radius ($1/e^2$) of the laser beam for local qubit control is set to $R_{\rm control}=1~\upmu \mathrm{m}$. The parameters for the trap and atom motions are summarized in Table \ref{tab:parameters}, which are commonly found in neutral-atom experiments~\cite{intro_global_bluvstein_quantum_2022, optical_misalign_chew_ultra-precise_2024, intro_scalability_pichard_rearrangement_2024}. 

\begin{table}
  \centering
  \setlength{\tabcolsep}{12pt}
  \renewcommand{\arraystretch}{1.5}
  \begin{tabular}{lr}
    \hline \hline
    Tweezer Beam Radius 
    ($1/e^2$) & $0.7~\upmu\mathrm{m}$ \\
    Trap Depth & $0.8~{\rm mK}$ \\
    Trap Frequency $\omega_r$ $(\omega_z)$& $ 2\pi\times155$ $(42)~{\rm kHz}$ \\
    \hline
    Atom Temperature & $30~{\rm \upmu K}$ \\
    Atom Distribution $\sigma_r~(\sigma_z)$ & $78~(201)~{\rm nm}$\\
    \hline \hline
  \end{tabular}
  \caption{
  \textbf{Parameters for the Trap and Atom Motions}  $\sigma_r~(\sigma_z)$ denotes the standard deviation of the atom's position along the radial (axial) direction.
  }
  \label{tab:parameters}
\end{table}

\begin{figure}[b]
  \centering
  \includegraphics[width=0.475\textwidth]{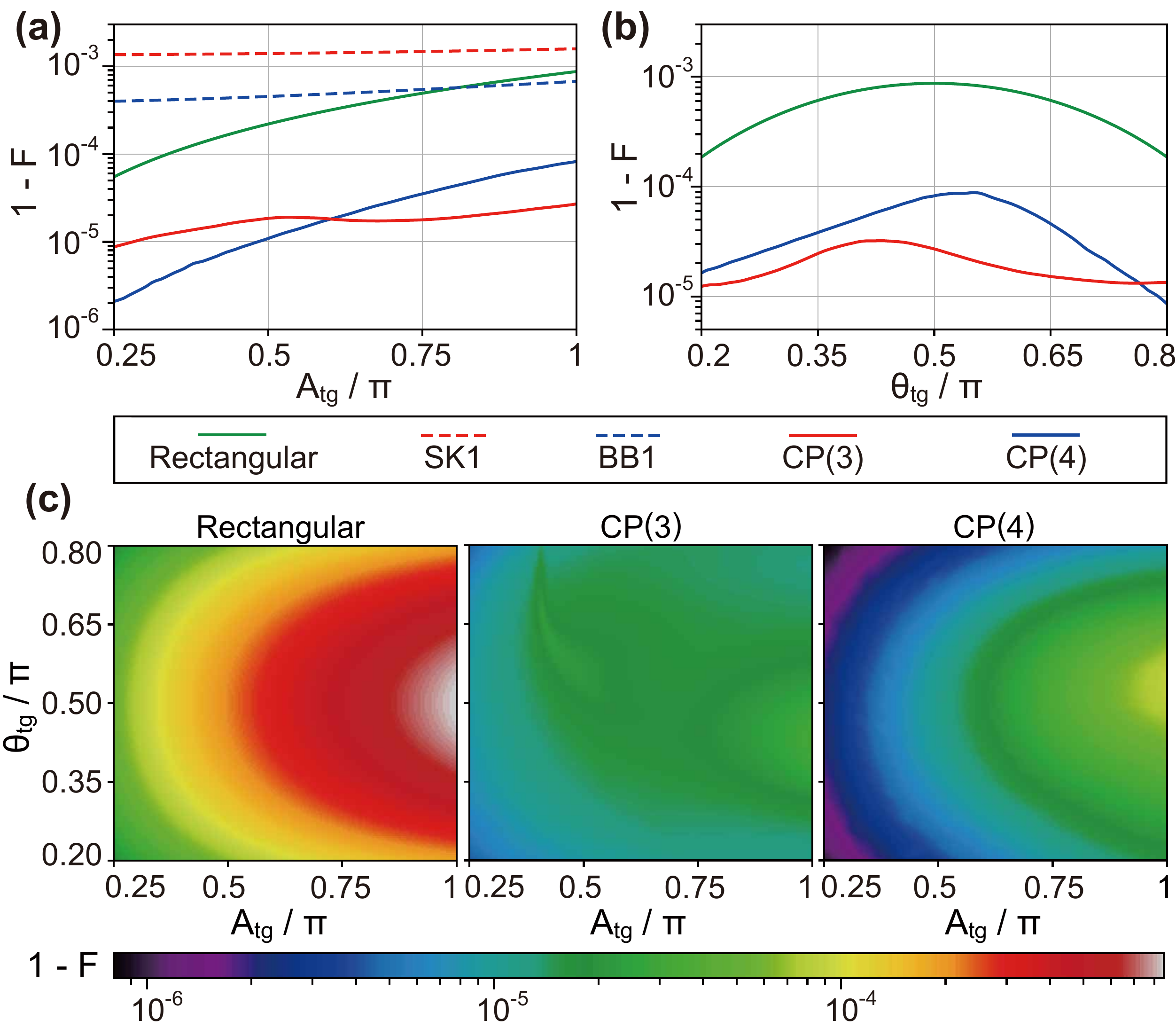}
  \caption{
  \textbf{Control Fidelities of Trained Composite Pulses}
      \textbf{(a)} and \textbf{(b)} Fidelity of CPs implementing target rotations with $\theta_\text{tg}=\pi/2$ and $A_\text{tg}=\pi$, respectively. 
      \textbf{(c)} Control fidelities over the training range.
    }
  \label{fig:fidelity}
\end{figure}

We specifically train the DNN for $n=3$ and $n=4$ so that the resulting sequences can be compared with the conventional CPs, \textit{SK}1~\cite{intro_CP_brown_arbitrarily_2004} and \textit{BB}1~\cite{intro_CP_wimperis_broadband_1994}, respectively, each of which uses the same number of pulses. Figures \ref{fig:trained-CPs}(b) and (c) show the pulse parameters of \textit{CP}(3) and \textit{CP}(4), respectively, obtained from two independently trained DNNs. Unlike \textit{SK}1 and \textit{BB}1, which are restricted to rotations with $\theta_\text{tg}=\pi/2$, the trained CPs can accommodate arbitrary target rotation axes within the training range $A_{\rm tg} \in [\pi/4, \pi]$ and $\theta_{\rm tg} \in [\pi/5, 4\pi/5]$. To ensure experimental feasibility, we constrain the pulse parameters such that $|\Omega_\text{c}|,|\Delta| \leq 2\pi\times 1~\text{MHz}$ (see Supp. F for details on the training precess).

Figure~\ref{fig:fidelity} plots the control fidelity of \textit{CP}($n$). As shown in Fig.~\ref{fig:fidelity}(a), the conventional CPs are ineffective at mitigating $\epsilon(t)$, exhibiting higher errors than a single rectangular pulse. By contrast, both AI-trained \textit{CP}(3) and \textit{CP}(4) coherently suppress this error, yielding an order-of-magnitude improvement in fidelity across a wide range of the qubit rotation (Figs.~\ref{fig:fidelity}(b) and (c)). For this assessment, we generated an independent validation set of 10,000 thermal atoms whose initial conditions were not used during the training.

This study applies directly to the local control of the microwave clock-state qubits (e.g., $\ket{0}=\ket{F=1,m_F =0}$ and $\ket{1}=\ket{F=2,m_F =0}$ of $^{87}$Rb) via the two-photon Raman transitions~\cite{intro_processor_bluvstein_logical_2024}, since the other time-varying effects such as the detuning error (due to the differential light shift~\cite{DRL_budget_kuhr_analysis_2005, DRL_budget_wang_polarizabilities_2016}), the qubit loss from the polarization mixing~\cite{DRL_budget_joseph_braat_imaging_2019}, and incoherent scattering~\cite{intro_fidel_levine_dispersive_2022} are negligible compared to the motion-induced amplitude error (Supp. I)

\section{Fidelity under Optical System Imperfections}
\begin{figure}[b]
  \centering
  \includegraphics[width=0.475\textwidth]{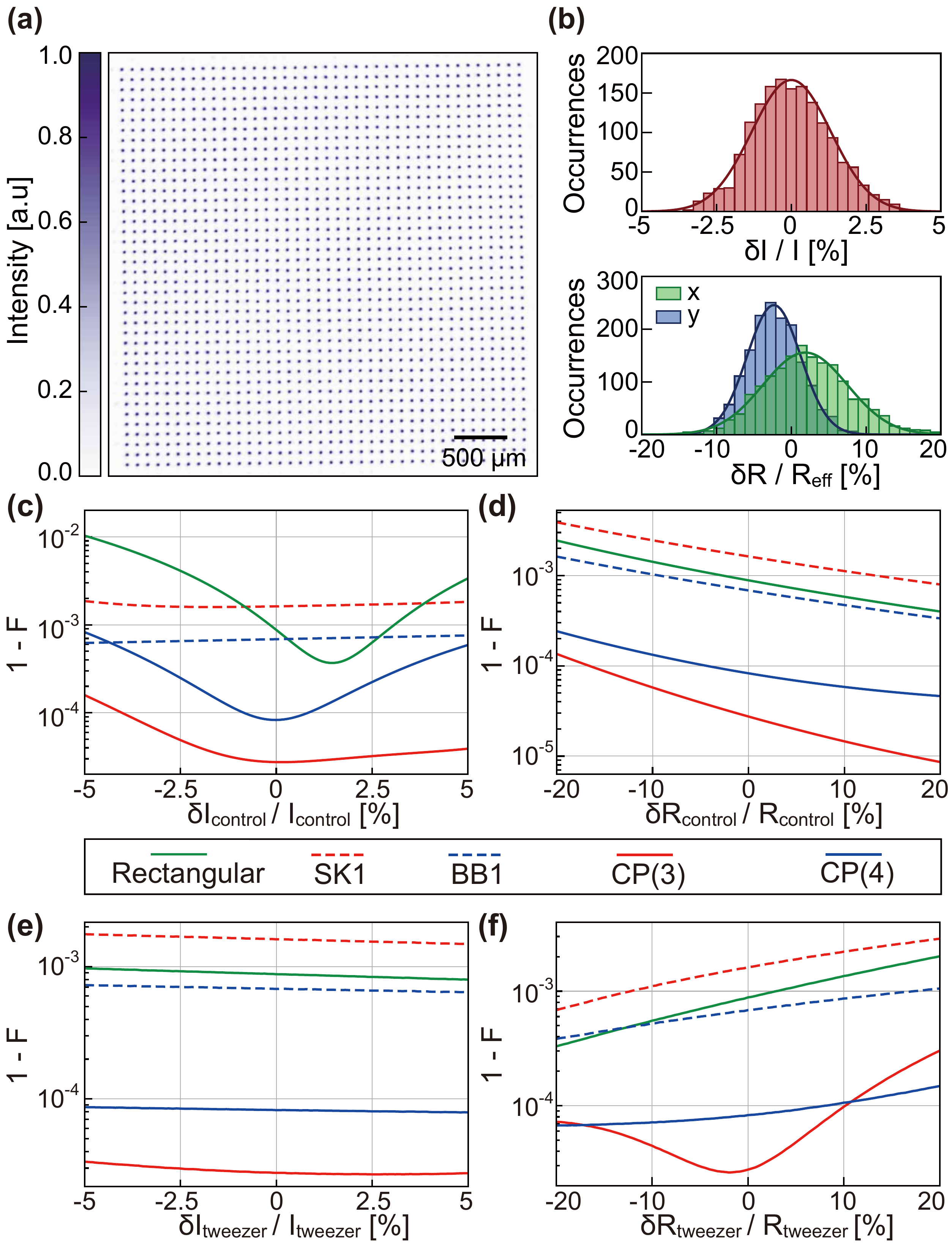}
  \caption{
    \textbf{Control Fidelities under Optical Focus Array}
    \textbf{(a)} Camera image of a $40\times40$ focus array generated by a phase SLM.
    \textbf{(b)} Distributions of peak intensity (top) and $1/e^2$ radius (bottom); solid lines are Gaussian fits with a standard deviation of 1.3\% for the intensity and 6.8\% (5.1\%) for the radius along $x$ ($y$) axis. $R_\text{eff}=\sqrt{R_xR_y}$ where $R_x$ ($R_y$) is the average radius along $x$($y$) axis.
    \textbf{(c)-(f)} $\pi$-pulse fidelities plotted against deviations in peak intensity and radius for the control beam \textbf{(c, d)} and tweezer \textbf{(e, f)}.
    }
  \label{fig:aberration}
\end{figure}
Gate implementations using tightly-focused control beams are inherently more sensitive to optical-system imperfections than schemes relying on global illumination. Residual aberrations~\cite{optical_imperfect_zhang_scaled_2024} and misalignments~\cite{optical_misalign_chew_ultra-precise_2024} between the control beam and the tweezer alter the distribution of motion-induced errors, thereby degrading the fidelity of CPs trained under ideal conditions. 

To quantify these effects, we generate a $40\times40$ optical focus (Fig.~\ref{fig:aberration}(a)) array with phase-only digital holography (see Supp. A for details)~\cite{intro_focus_kim_large-scale_2019} -- a core technique for both the generation~\cite{intro_barredo_atom-by-atom_2016} and control~\cite{optical_imperfect_de_oliveira_demonstration_2025, intro_processor_manovitz_quantum_2025} of neutral-atom qubit arrays. After adaptive aberration correction and intensity homogenization, residual variations in peak intensity and $1/e^2$-radius persist across the array (Fig.~\ref{fig:aberration}(b)). We treat these distributions as realistic inhomogeneities in the intensity and radius of both control and tweezer beams and evaluate the fidelity accordingly. 

When evaluated over the measured residual aberrations, \textit{CP}(3) and \textit{CP}(4) outperform both conventional CPs and rectangular pulse (Figs.~\ref{fig:aberration}(c)-(f)). The fidelity dependence on the control-beam inhomogeneity reflects the spatial landscape of Rabi frequency experienced by atoms as they traverse the beam. Although the trained CPs remain somewhat sensitive to the intensity variation $\delta I_\text{control}$ of the control beam (Fig.~\ref{fig:aberration}(c)), they still achieve markedly higher fidelities than conventional robust CPs. Because a larger control beam radius mitigates motion-induced error~\cite{intro_processor_bluvstein_logical_2024}, fidelity increases monotonically with beam radius (Fig.~\ref{fig:aberration}(d)). 

\begin{figure}[b]
  \centering
  \includegraphics[width=0.475\textwidth]{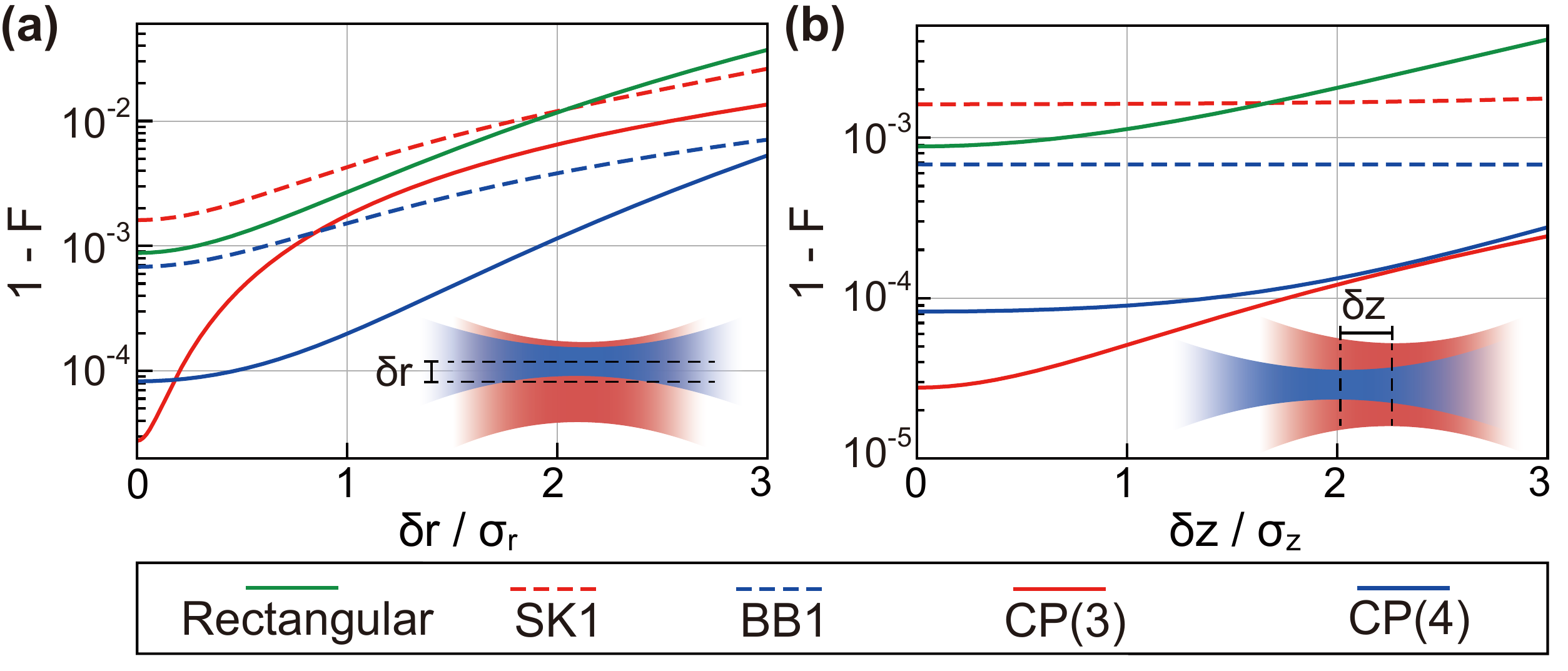}
  \caption{
    \textbf{Control Fidelities under Misalignments}
    \textbf{(a)} and \textbf{(b)} $\pi$-pulse fidelities plotted against misalignment in the radial and axial directions, respectively. Insets show the control beam (red) offset from the tweezer (blue).
    }
  \label{fig:misalign}
\end{figure}

Meanwhile, inhomogeneities in the tweezer beams modify the trap depth and frequency, changing both the range and characteristic frequency of the atom motion in the tweezer (Supp. D). As shown in Figs.~\ref{fig:aberration}(e) and (f), a narrower and deeper trap enhances the control fidelity due to a tighter atom localization. The \textit{CP}(3) shows a notable dependence on $\delta R_\text{tweezer}$, suggesting it has been trained to suppress specific trap frequencies. The spectral properties of the trained CPs are analyzed in the following section.

We next evaluate how relative displacements between the control beam and the tweezer reduce the control fidelity. Misalignments modify the local Rabi frequency experienced by the atom: rather than the ideal quadratic dependence, the frequency exhibits a linear variation with displacement. Consequently, fidelity decreases, as shown in Figs.~\ref{fig:misalign}(a) and (b); radial offsets produce a much larger loss than axial ones. Notably, \textit{CP}(3) quickly loses its error-mitigation capability, whereas \textit{CP}(4) remains relatively robust over the misalignment. The contrast arises from the distinct error-filtering characteristics of the two trained CPs with the additional spectral components in $\epsilon(t)$ due to the misalignments, as analyzed in the next section.

\section{Spectral Analysis}

\begin{figure*}
  \centering
  \includegraphics[width=1\textwidth]{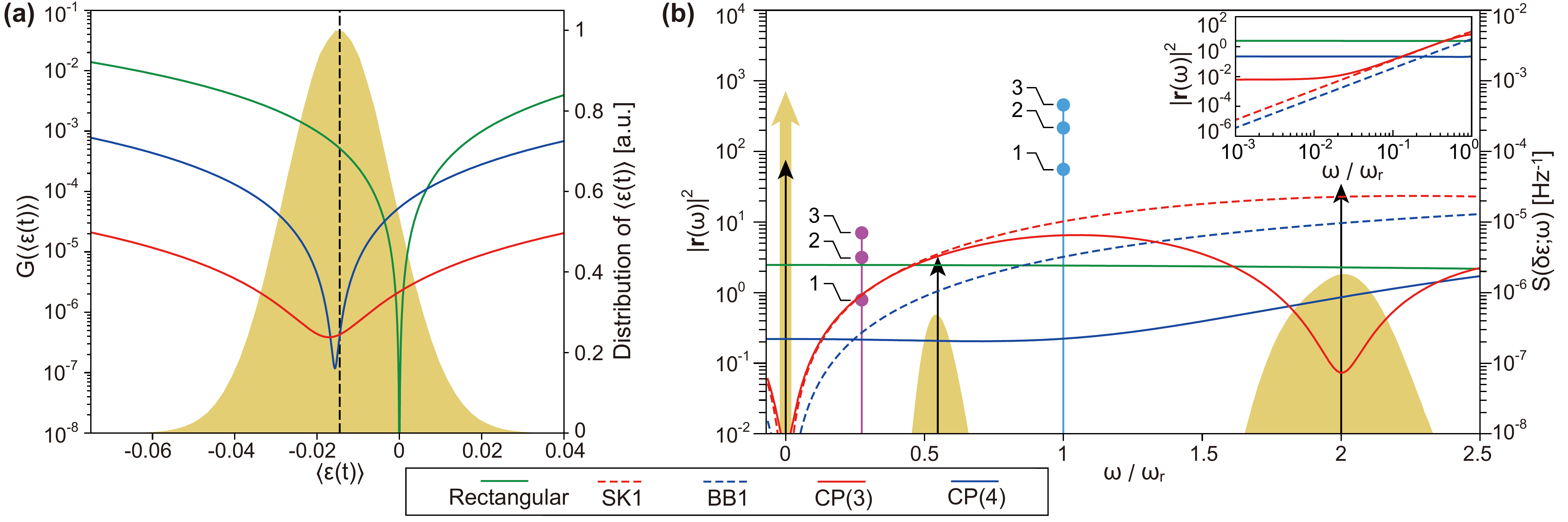}
  \caption{
    \textbf{Spectral Characteristics}
    \textbf{(a)} The residual bias $G(\langle\epsilon(t)\rangle)$ of the trained CPs for a $\pi$-pulse, overlaid with the error bias $\langle\epsilon(t)\rangle$ for an ideal focus (dashed black line) and generated foci in Fig. \ref{fig:aberration} (a) (yellow). $G(\langle\epsilon(t)\rangle)=0$ for \textit{SK}1 and \textit{BB}1 (not shown).
    \textbf{(b)} Filter functions of the CPs, overlaid with the power spectra of the ideal focus (black) and generated foci (yellow). The cyan (magenta) points denote spectral components due to radial (axial) misalignments. Numbers labeling the points indicate the magnitude of misalignment in units of $\delta r/\sigma_r$ for radial and $\delta z/\sigma_z$ for axial displacements. Inset: zoom near the DC component.
    }
  \label{fig:spectral}
\end{figure*}

In this section, we analyze the spectral characteristics of the trained CPs that suppress the motion-induced amplitude error $\epsilon(t)$. In the composite-pulse frame defined by $U_\text{c}$ (Supp. C), we show that to the leading order in $\epsilon(t)$
\begin{equation*}
    1-F \simeq G(\langle\epsilon(t)\rangle) + \frac{1}{2\pi}\int d\omega |\boldsymbol{r}(\omega)|^2 S(\delta\epsilon;\omega),
\end{equation*}
where the two contributions stem from (i) imperfect cancellation of the mean amplitude error $\langle\epsilon(t)\rangle$ (i.e., bias)
\begin{equation*}
G(\langle\epsilon(t)\rangle)=|\langle\epsilon(t)\rangle\boldsymbol{r}(0) - \boldsymbol{\mathcal{D}}|^2,
\end{equation*}
and (ii) incomplete filtering of the zero-mean fluctuation $\delta\epsilon \doteq \epsilon(t)-\langle\epsilon(t)\rangle$ characterized by its power spectrum $S(\delta\epsilon;\omega)$. Here, $\boldsymbol{\mathcal{D}}=\frac{1}{2}\text{Im}(\text{Tr}[\boldsymbol{\sigma}\log(\mathcal{U}^\dagger U_\text{c})])$ with the Pauli matrices $\boldsymbol{\sigma}=(\hat\sigma_x,\hat\sigma_y,\hat\sigma_z)$, and 
\begin{equation*}
    \boldsymbol{r}(\omega) = \frac{1}{2\hbar}\int_{-\infty}^{\infty}\text{Tr}[\boldsymbol{\sigma}\tilde{H}_\text{c}^\perp(t)] e^{-i\omega t}dt
\end{equation*}
is the filter amplitude obtained from the transverse CP Hamiltonian $\tilde{H}_\text{c}^\perp(t)$ (Supp. B). Note that $G(\langle\epsilon(t)\rangle)$ captures the residual error bias, while $\boldsymbol{r}(\omega)$ characterizes the CP's spectral response. 

Figures \ref{fig:spectral}(a) and (b) plot $G(\langle\epsilon(t)\rangle)$ and $|\boldsymbol{r}(\omega)|^2$, respectively, overlaid with the experimentally inferred distributions of $\langle\epsilon(t)\rangle$ and $S(\delta\epsilon;\omega)$. These distributions are obtained from the measured intensity and beam-radius variations across the 40 $\times$ 40 focus array shown in Fig.~\ref{fig:aberration}(a). For a rectangular pulse, the residual bias $G(\langle\epsilon(t)\rangle)$ reaches its minimum at zero mean error, $\langle\epsilon(t)\rangle = 0$. By contrast, the trained CPs are optimized to minimize $G(\langle\epsilon(t)\rangle)$ at the mean error due to the atom's harmonic motion (black dotted line in Fig.~\ref{fig:spectral}(a)). This designed shift effectively displaces the implemented rotation $U_\text{c}$ from the target gate $\mathcal{U}$, such that $\boldsymbol{\mathcal{D}} \simeq \langle\epsilon(t)\rangle\boldsymbol{r}(0)$, thereby canceling the leading-order error bias. Moreover, the broader width of $G(\langle\epsilon(t)\rangle)$ for the trained CPs indicates enhanced robustness against variations in control-beam intensity $\delta I_\text{control}$ and alignment imperfections that modify the spatial intensity profile.

The filter functions $|\boldsymbol{r}(\omega)|^2$ of the trained CPs are tailored to suppress the dominant AC components in $S(\delta\epsilon;\omega)$, particularly at $\omega = 2\omega_r$ and $2\omega_z$. These components arise from the atom motions in the control beam's quadratic Rabi-frequency envelope. While both trained CPs exhibit saturated filtering near DC (see inset of Fig.~\ref{fig:spectral}(b)), their effectiveness in suppressing motion-induced fluctuations hinges on the targeted reduction of these harmonic components. Notably, the two trained CPs demonstrate distinct spectral characteristics. The \textit{CP}(3) features narrowband filtering centered at the dominant $2\omega_r$. By contrast, \textit{CP}(4) exhibits broadband filtering that extends over all relevant spectral components, including $\omega = 2\omega_z$. This broadband profile results in resilience under alignment errors (Fig.~\ref{fig:misalign}), which introduce additional spectral weight at $\omega = \omega_r$ and $\omega_z$ (cyan and magenta points in Fig.~\ref{fig:spectral}(b), respectively) due to the linear Rabi-frequency gradient sampled by off-centered atoms. Similar spectral responses are observed for the trained CPs implementing different target rotations (Supp Fig. S7).

\section{Conclusion}

We demonstrate that AI-trained composite pulses enable precise and efficient single-qubit control with spatially localized laser fields. By employing deep reinforcement learning to model the interaction dynamics of atoms oscillating in optical tweezers, the trained CPs effectively suppress motion-induced amplitude errors that arise from the non-uniform intensity profile of the control beam. We show that the resulting localized control yields gate fidelities exceeding those of analytic sequences such as \textit{BB}1 and \textit{SK}1 by an order of magnitude, even in the presence of optical aberrations and alignment imperfections.

In addition, the trained neural network exhibits generalization over the space of SU(2) rotations, enabling automated identification of composite pulses for arbitrary single-qubit gates. By substituting the CPs already employed in neutral-atom experiments~\cite{intro_processor_bluvstein_logical_2024, intro_local_muniz_high-fidelity_2025, intro_local_lis_midcircuit_2023}, our method provides a general pathway to high-precision qubit control and enlarges the capabilities of neutral-atom quantum processors. This ability to learn and generalize atom-laser interactions underscores a key advantage of AI-optimized control for neutral-atom qubits. Owing to its adaptability to complex, system-specific error landscapes, our approach is readily applicable to other quantum platforms, including trapped ions~\cite{conc_ion_leu_fast_2023, conc_ion_ringbauer_universal_2022} and solid-state qubits~\cite{conc_super_fang_nonadiabatic_2024, conc_super_tonchev_robust_2025, conc_semi_ni_diverse_2025, intro_optim_xie_9992-fidelity_2023}.

\bibliography{main}

% \section*{Funding}
% This research is supported in part by the KAIST UP Program and POSCO Science Fellowship.

% \section*{Author Contributions}
% S.P. and D.K. conceived the idea, developed the theoretical formalism, and wrote the manuscript. S.P. performed and analysed the numerical calculations with assistance from S.L. and M.K.. S.P. and K.L. carried out the experiments and analysed the experimental data. D.K. supervised the project. All authors discussed the results and contributed to the final manuscript.

% \section*{Competing Interest}
% This work was filed as a Korea Patent Application (No.10-2025-0051791). D.K. is the founder and CEO of OQT Inc., and is a shareholder of QuEra Computing Inc.

\end{document}

\title{Supplementary Material}
\maketitle

\makeatletter
\renewcommand{\theequation}{S\arabic{equation}}
\renewcommand{\thefigure}{S\arabic{figure}}
\renewcommand{\bibnumfmt}[1]{[S#1]}
\renewcommand{\citenumfont}[1]{S#1}

\subsection{Programmable Focus Array Generation}

\begin{figure}[htb!]
    \centering
    \includegraphics[width=1\linewidth]{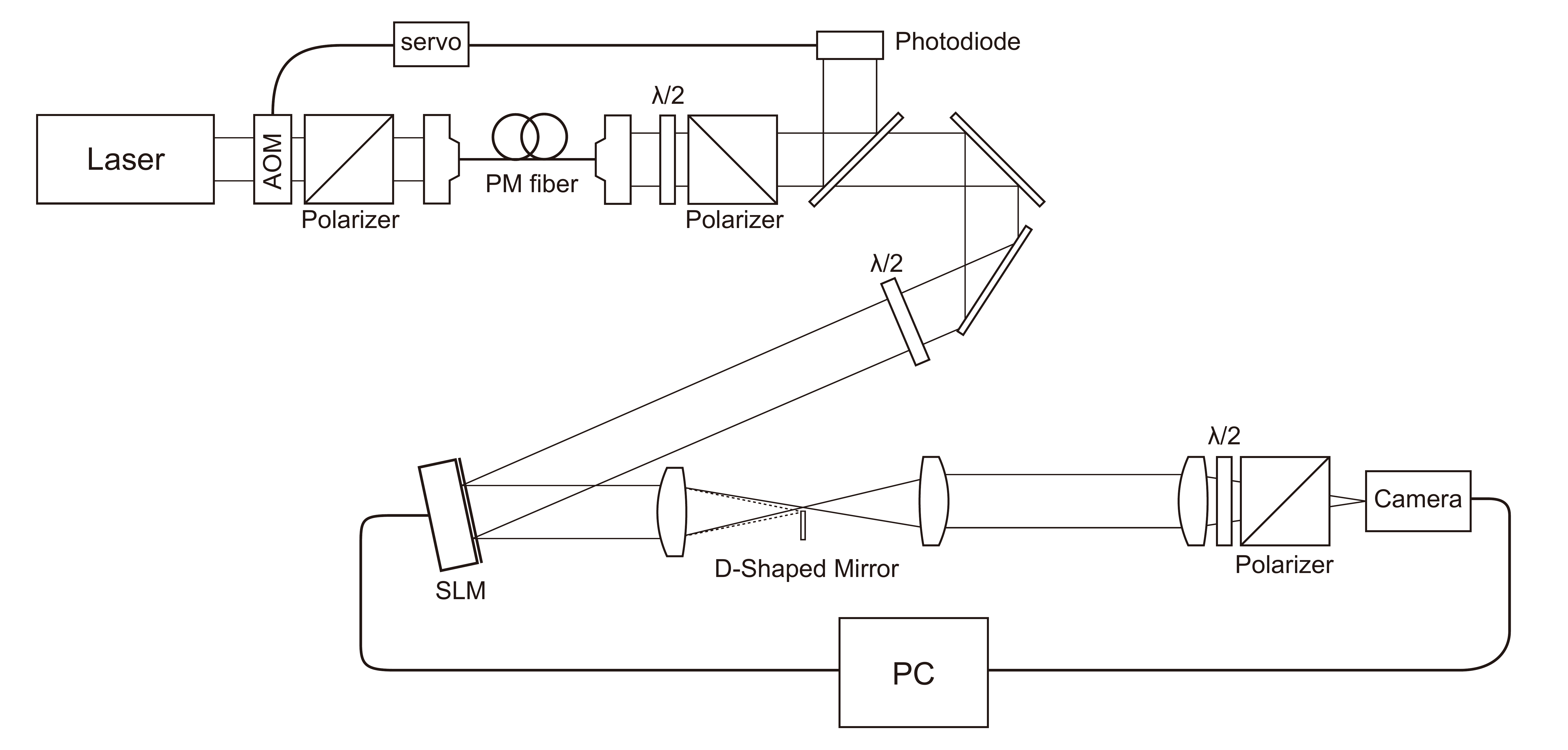}
    \caption{\textbf{Focus Array Generation Setup}}
    \label{fig:focus-array-setup}
\end{figure}

We employ a liquid-crystal-on-silicon spatial light modulator (SLM) to engineer the spatial profile of a laser beam, enabling the creation of both focused qubit control beams and optical tweezers. Initially, a 532~\text{nm} laser beam is spatially filtered through a polarization-maintaining single-mode fiber to ensure a high-quality mode before it illuminates the SLM. The unwanted zeroth-order diffraction is spatially filtered at the first focal plane using a D-shaped mirror. The phase hologram displayed on the SLM is calculated using a phase-fixed weighted Gerchberg-Saxton algorithm~\cite{intro_focus_kim_large-scale_2019, supp_KYL_matsumoto_high-quality_2012, supp_KYL_nogrette_single-atom_2014, supp_KYL_tamura_highly_2016}. This calculation incorporates adaptive correction for optical aberrations and is further refined using camera feedback for intensity homogenization across the array. The beam reflected from the SLM then forms a $40\times40$ array of focused optical spots, as shown in Fig.~4(a) in the main text.

\subsection{The Composite Pulse (CP) Frame}

The system Hamiltonian is given by:
\begin{align}
    H(t;\epsilon(t)) &= H_\textrm{c}(t) + H'(t;\epsilon(t)),
    \label{eq:system-hamiltonian}
\end{align}
which describes the interaction of an atom in an optical tweezer with a control laser field. $U(t;\epsilon(t))$ denotes the corresponding time evolution operator of $H(t;\epsilon(t))$

The first term, $H_\textrm{c}(t)$, represents the control Hamiltonian of the composite pulse (CP) sequence:
\begin{equation*}
    H_\textrm{c}(t)
    = \frac{\hbar}{2}\sum_{k=1}^n 
     W_k(t)
    \begin{bmatrix}
        \text{Re}(\Omega_{\textrm{c},k}) \\
        \text{Im}(\Omega_{\textrm{c},k}) \\
        \Delta_k
    \end{bmatrix}
    \cdot 
    \begin{bmatrix}
        \hat\sigma_x \\
        \hat\sigma_y \\
        \hat\sigma_z
    \end{bmatrix}
    = H_\text{c}^\perp(t) + H_\text{c}^\parallel(t)
\end{equation*}
with Pauli matrices $\boldsymbol{\sigma}= (\hat\sigma_x,\hat\sigma_y,\hat\sigma_z)$. Here, $H_\text{c}^\perp(t)$ is the transverse component of $H_\text{c}(t)$, proportional to $\hat\sigma_x$ and $\hat\sigma_y$, whereas $H_\text{c}^\parallel(t)$ is the longitudinal component, proportional to $\hat\sigma_z$.
$W_k(t)$ ($k=1,2,\dots,n$) is a unit rectangular function that defines a time window of the $k$-th pulse, which has a duration $\tau_k$ and starts at $T_{k}=\sum_{l=1}^{k-1}\tau_l$ (with $T_1=0$). The parameters $\Omega_{\textrm{c},k}$ and $\Delta_k$ denote the Rabi frequency and detuning of the $k$-th pulse, respectively. $U_\text{c}(t)$ denotes the time-evolution operator due to $H_\text{c}(t)$.

The second term,
\begin{equation*}
    H'(t;\epsilon(t))
    = \epsilon(t) H_\text{c}^\perp(t)
\end{equation*}
accounts for the time-dependent perturbation on $H_\textrm{c}(t)$ due to the motion-induced amplitude error $\epsilon(t)$.
Since we consider amplitude error only, $H'(t;\epsilon(t))$ excludes the $\sigma_z$ component, which is associated with detuning error.

Within the composite-pulse (CP) frame defined by $U_\text{c}(t)$ such that $|\tilde\psi(t;\epsilon(t))\rangle\doteq U^\dagger_\text{c}(t)|\psi(t;\epsilon(t))\rangle$, 
\begin{align*}
i\hbar\frac{d}{dt}|\tilde{\psi}(t;\epsilon(t))\rangle &= i\hbar\frac{d}{dt}U^\dagger_\text{c}(t)\ket{\psi(t;\epsilon(t))} \\
&=i\hbar\left(U_\text{c}^\dagger(t)\frac{d\ket{\psi(t;\epsilon(t))}}{dt}+\frac{dU_\text{c}^\dagger(t)}{dt}\ket{\psi(t;\epsilon(t))}\right) \\
&=U_\text{c}^\dagger(t)\big(H(t;\epsilon(t))-H_\text{c}(t)\big)\ket{\psi(t;\epsilon(t))} \\
&=U^\dagger_\text{c}(t)H'(t;\epsilon(t))U_\text{c}(t)|\tilde{\psi}(t;\epsilon(t))\rangle\\
&=\epsilon(t)\tilde H^\perp_\text{c}(t)|\tilde{\psi}(t;\epsilon(t))\rangle,
\end{align*}
where $\tilde H^\perp_\text{c}(t)=U^\dagger_\text{c}(t)H^\perp_\text{c}(t)U_\text{c}(t)$ is the transverse CP Hamiltonian in the CP frame. We define the time-evolution operator due to $\epsilon(t)\tilde H_\text{c}^\perp(t)$ as $\tilde U(t;\epsilon(t))$, describing the evolution due to $\epsilon(t)$.
By using the Magnus expansion~\cite{spectral_response_green_arbitrary_2013}, we find that 
\begin{equation*}
    \tilde{U}(T;\epsilon(t)) \doteq \exp(-i \boldsymbol{a}(\epsilon(t))\cdot\boldsymbol{\sigma}),
\end{equation*}
where
\begin{align}
    \boldsymbol{a}(\epsilon(t))
    &=\frac{1}{2\hbar}\int^T_0 dt_1\epsilon(t_1)\text{Tr}[\boldsymbol{\sigma} \tilde{H}^\perp_\text{c}(t_1)]
    +\left(\frac{1}{2\hbar}\right)^2\int^T_0 dt_1\int^{t_1}_0 dt_2\epsilon(t_1)\epsilon(t_2)\text{Tr}[\boldsymbol{\sigma} \tilde{H}^\perp_\text{c}(t_1)]\cross\text{Tr}[\boldsymbol{\sigma} \tilde{H}^\perp_\text{c}(t_2)]
    +\dots \notag\\
    &\simeq\frac{1}{2\hbar}\int^T_0 dt_1\epsilon(t_1)\text{Tr}[\boldsymbol{\sigma} \tilde{H}^\perp_\text{c}(t_1)].
    \label{eq:vector_a}
\end{align}

\subsection{The Derivation of Spectral Analysis}
The control fidelity $F$ of implementing target rotation $\mathcal{U}$ can be written in terms of (\ref{eq:vector_a}):
\begin{align}
    F
    &\doteq \frac{1}{4}\langle|\text{Tr}[\mathcal{U}^\dagger U(T;\epsilon(t))]|^2\rangle \notag\\
    &= \frac{1}{4}\langle|\text{Tr}[\mathcal{U}^\dagger U_\mathrm{c}(T)\tilde{U}(T;\epsilon(t))]|^2\rangle \notag\\
    &= \frac{1}{4}\langle|\text{Tr}[\exp(i\boldsymbol{\mathcal{D}}\cdot\boldsymbol{\sigma})\exp(-i\boldsymbol{a}(\epsilon(t))\cdot\boldsymbol{\sigma})]|^2\rangle \notag\\
    &= \langle|\cos(|\boldsymbol{a}(\epsilon(t))|)\cos(|\boldsymbol{\mathcal{D}}|)+\frac{\boldsymbol{a}(\epsilon(t))\cdot\boldsymbol{\mathcal{D}}}{|\boldsymbol{a}(\epsilon(t))||\boldsymbol{\mathcal{D}}|}\sin(|\boldsymbol{a}(\epsilon(t))|)\sin(|\boldsymbol{\mathcal{D}}|)|^2\rangle \notag\\
    &\simeq 1 - \langle|\boldsymbol{a}(\epsilon(t)) - \boldsymbol{\mathcal{D}}|^2\rangle,
    \label{eq:control-fidelity}
\end{align}
where $\langle\dots\rangle$ denotes an average over the initial position and velocity of the thermal ensemble of trapped atoms, 
and 
\begin{equation*}
    \boldsymbol{\mathcal{D}}=\frac{1}{2}\text{Im}[\text{Tr}[\boldsymbol{\sigma}\log(\mathcal{U}^\dagger U_\textrm{c}(T))]].
\end{equation*}
Here, we use $U(t;\epsilon(t))=U_\text{c}(t)\tilde{U}(t;\epsilon(t))$ as shown in
\begin{equation*}
    U(t;\epsilon(t))|\psi(0)\rangle = |\psi(t;\epsilon(t))\rangle=U_\text{c}(t)|\tilde\psi(t;\epsilon(t))\rangle = U_\text{c}(t)\tilde{U}(t;\epsilon(t))|\psi(0)\rangle. 
\end{equation*}

By decomposing $\epsilon(t)$ into a mean amplitude error $\langle\epsilon(t)\rangle$ and zero-mean fluctuation $\delta\epsilon(t)\doteq\epsilon(t)-\langle\epsilon(t)\rangle$, the residual infidelity (\ref{eq:control-fidelity}) can be expressed as distinct components:
\begin{align}
        1 - F
        &= \langle|\boldsymbol{a}(\langle\epsilon(t)\rangle) + \boldsymbol{a}(\delta\epsilon(t))|^2\rangle - 2\boldsymbol{a}(\langle\epsilon(t)\rangle)\cdot\boldsymbol{\mathcal{D}} + |\boldsymbol{\mathcal{D}}|^2 \notag\\
        &= |\boldsymbol{a}(\langle\epsilon(t)\rangle) - \boldsymbol{\mathcal{D}}|^2 + \langle|\boldsymbol{a}(\delta\epsilon(t))|^2\rangle
        \label{eq:residual-infidelity}
\end{align}

To the leading order of $\epsilon(t)$, we find that 
\begin{align*}
    |\boldsymbol{a}(\langle\epsilon(t)\rangle) - \boldsymbol{\mathcal{D}}|^2
    &=\left|\frac{1}{2\hbar}\int^T_0\langle\epsilon(t)\rangle\text{Tr}[\boldsymbol{\sigma}\tilde{H}^\perp_\text{c}(t)]dt-\frac{1}{2}\text{Im}[\text{Tr}[\boldsymbol{\sigma}\log(\mathcal{U}^\dagger U_\text{c}(T))]]\right|^2\\
&=\left|\langle\epsilon(t)\rangle\boldsymbol{r}(0)-\boldsymbol{\mathcal{D}}\right|^2 \doteq G(\langle\epsilon(t)\rangle),
\end{align*}
where
\begin{align*}
    \boldsymbol{r}(\omega) &\doteq \frac{1}{2\hbar}\int^\infty_{-\infty}\text{Tr}[\boldsymbol{\sigma}\tilde{H}^\perp_\text{c}(t)]e^{-i\omega t}dt\\
    &=\frac{1}{2\hbar}\mathcal{F}\{\text{Tr}[\boldsymbol{\sigma}\tilde{H}^\perp_\text{c}(t)]\}(\omega) 
\end{align*}
is the filter amplitude. Here, $\mathcal{F}\{.\}$ denotes the Fourier transform. 

For the $\delta\epsilon(t)$ contribution in (\ref{eq:residual-infidelity}), 
\begin{align*}
    \langle|\boldsymbol{a}(\delta\epsilon(t))|^2\rangle
    &\simeq\left(\frac{1}{2\hbar}\right)^2\int_0^{T} dt_1 \int_0^{T} dt_2 \langle\delta\epsilon(t_1) \delta\epsilon(t_2)\rangle\text{Tr}[\boldsymbol{\sigma}\tilde{H}^\perp_\text{c}(t_1)]\cdot\text{Tr}[\boldsymbol{\sigma}\tilde{H}^\perp_\text{c}(t_2)]\\
    &= \left(\frac{1}{2\hbar}\right)^2\int_0^{T} dt_1 \int_0^{T} dt_2 \left(\frac{1}{2\pi}\int^\infty_{-\infty}d\omega S(\delta\epsilon;\omega) e^{i\omega (t_2 - t_1)}\right)\text{Tr}[\boldsymbol{\sigma}\tilde{H}^\perp_\text{c}(t_1)]\cdot\text{Tr}[\boldsymbol{\sigma}\tilde{H}^\perp_\text{c}(t_2)]\\
    &= \frac{1}{2\pi} \int^\infty_{-\infty}d\omega S(\delta\epsilon;\omega) |\boldsymbol{r}(\omega)|^2.
\end{align*}
Here, we have used the wide-sense stationarity of $\delta\epsilon(t)$, which is valid for thermal atoms whose oscillatory motion has a uniformly distributed phase. Under this condition, the auto-correlation depends only on the time difference $t_2-t_1$, and the power spectral density of $\delta\epsilon(t)$ is
\begin{align*}
S(\delta\epsilon;\omega) &= \int_{-\infty}^{\infty}  \langle\delta\epsilon(t_1)\delta\epsilon(t_2)\rangle e^{-i\omega (t_2-t_1)} d(t_2-t_1)\\
    &=|\mathcal{F}\{\delta\epsilon(t)\}(\omega)|^2, 
\end{align*}
and its filter is given by $|\boldsymbol{r}(\omega)|^2$.

\subsection{Motion Induced Amplitude Error\label{section:motion-induced-amplitude-error}}
To calculate the motion induced amplitude error, we assume that the control beams and tweezers follow an elliptical Gaussian beam with the intensity profile~\cite{supp_arnaud_gaussian_1969}:
\begin{align*}
    I(x,y,z) &= I_0u_x(x,z)u_y(y,z)
\end{align*}
where
\begin{align*}
    u_x(x,z) = \sqrt{\frac{z_x^2}{z^2+z_x^2}}\exp(-2\frac{x^2}{R_x^2}\frac{z_x^2}{z^2+z_x^2}),\qquad u_y(y,z) = \sqrt{\frac{z_y^2}{z^2+z_y^2}}\exp(-2\frac{y^2}{R_y^2}\frac{z_y^2}{z^2+z_y^2}).
\end{align*}
Here, $I_0$ is peak intensity, $R_i$ and $z_i$ are $1/e^2$ radius and Rayleigh range along the $i=x,y$ components. Near the maximum, the intensity can be approximated by a quadratic curvature:
\begin{align}
    I(x,y,z) 
    &\simeq I_0 \left(1-2\frac{x^2}{R_x^2} - 2\frac{y^2}{R_y^2} - \frac{z^2}{z_0^2}\right),
    \label{eq:approx-intensity}
\end{align}
with an effective Rayleigh range:
\begin{equation*}
    \frac{1}{z_0^2} \doteq \frac{1}{2}\left(\frac{1}{z_x^2}+\frac{1}{z_y^2}\right).
\end{equation*}

Based on (\ref{eq:approx-intensity}), we determine the trap depth:
\begin{align*}
    V_0 = \alpha I_{\text{tweezer},0},
\end{align*}
and trap frequencies:
\begin{align*}
    \omega_x^2 = \frac{4V_0}{m_\text{atom}R_{\text{tweezer}, x}^2},~~~\omega_y^2 = \frac{4V_0}{m_\text{atom}R_{\text{tweezer},y}^2},~~~\omega_z^2 = \frac{2V_0}{m_\text{atom}z_{\text{tweezer},0}^2}.
\end{align*}
where $m_\text{atom}$ is atomic mass and $\alpha$ is the dynamical polarizability of the $5^2S_{1/2}$ manifold of ${}^{87}\text{Rb}$ for the wavelength of the optical tweezer~\cite{supp_pol_grimm_optical_2000}.

The initial condition of a trapped atom is sampled from a Gaussian distribution $\mathcal{N}(\mu,\sigma^2)$ with mean $\mu$ and standard deviation $\sigma$:
\begin{equation}
 x_i(t=0), \frac{v_i(t=0)}{\omega_i} \sim \mathcal{N}(0, k_B T_\text{atom} / m \omega_i^2)
 \label{eq:initial-condition}
\end{equation}
where $T_\text{atom}$ is the temperature of trapped atom and $i = x,y,z$. The distribution is derived from the partition function of the canonical ensemble of a harmonic oscillator at temperature $T_\text{atom}$. The atom motion is treated classically since the thermal energy of the atom is greater than the quantization energy of the harmonic potential $k_B T_\text{atom} \gg \hbar\omega_i$, and the harmonic approximation is applied to the trap potential since the trap depth is greater than thermal energy $V_0 \gg k_B T_\text{atom}$~\cite{intro_motion_tuchendler_energy_2008}.

From the initial conditions, the trajectory of the trapped atom is
\begin{equation*}
    x_i(t) = X_i\sin(\omega_i t + \Phi_i),
\end{equation*}
where
\begin{equation*}
    X_i = \sqrt{x^2_i(0) + v^2_i(0)/\omega^2_i},
    \qquad
    \Phi_i = \arctan(\frac{\omega_i x_i(0)}{v_i(0)}).
\end{equation*}

Along the trajectory, the time-varying Rabi frequency $\Omega(\boldsymbol{x}(t))$ is sampled to derive the motion-induced error:
\begin{align*}
    \epsilon(t)
    &= \Omega(\boldsymbol{x}(t))/\Omega_\text{c} - 1\\
    &= I_\text{control}(\boldsymbol{x}(t))/I_{\text{control},0} - 1\\
    &\simeq -\sum_{i=x,y}\frac{X_i^2}{R_{\text{control},i}^2}(1+\cos(2(\omega_i t + \Phi_i))) - \frac{X_z^2}{2 z_{\text{control},0}^2} (1+\cos(2(\omega_z t + \Phi_z))).
\end{align*}
Here, we consider the two-photon Raman transitions between $|F=1, m_F=0\rangle$ and $|F=2, m_F=0\rangle$ states in $\rm 5^2 S_{1/2}$ manifold of $^{87}$Rb. Therefore, the Rabi frequency $\Omega(\boldsymbol{x})$ is proportional to the intensity of the control beam $I_\text{control}(\boldsymbol{x})$. To calculate the control fidelity, we explicitly evaluate the motion-induced error without approximation. However, in its approximated form, we find that the error consists of a static component, originating from the offset of the motion from the center, and $\omega=2\omega_r$ and $2\omega_z$ components arising from the different trap frequencies along each axis. Note that, in the presence of misalignment between tweezer and control beam, the approximation in (\ref{eq:approx-intensity}) introduces linear terms which produce the $\omega=\omega_r$ and $\omega_z$ components in $\epsilon(t)$.

\subsection{Fidelity Calculation}
\begin{figure*}
  \centering
  \includegraphics[width=1\textwidth]{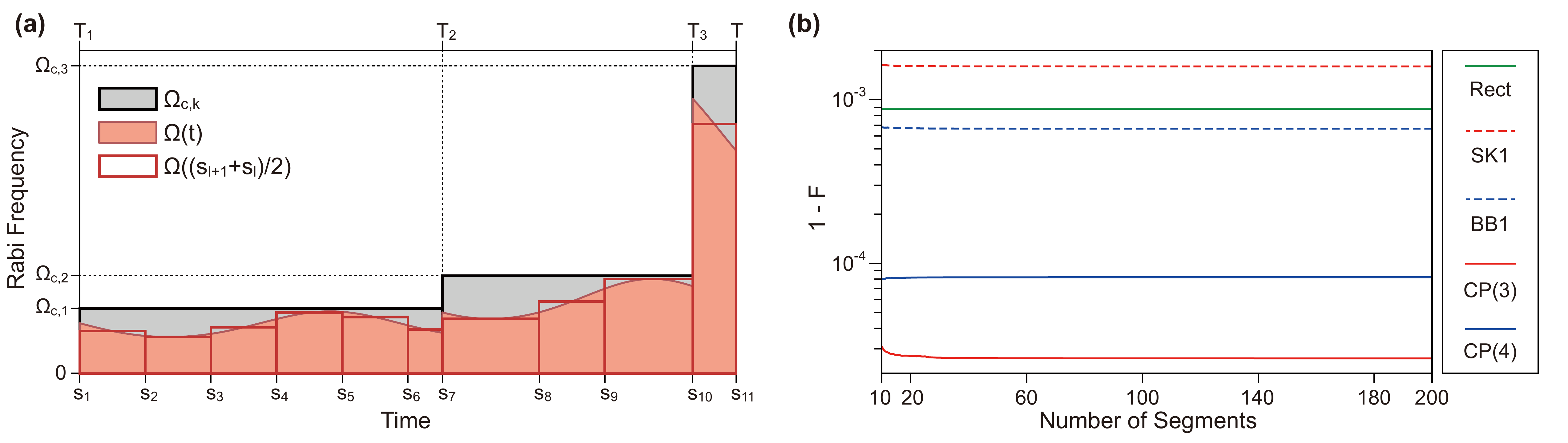}
  \caption{
    \textbf{Fidelity Calculation}
    \textbf{(a)} The time-varying Rabi frequency and its discretization into a total of $m=10$ segments. Note that, the segment boundaries $s_7$ and $s_{10}$ are adjusted to align with $T_2$ and $T_3$.
    \textbf{(b)} Control fidelity with respect to the number of segments for CPs encoding a $\pi$-pulse.
        }
  \label{fig:fidelity-calc}
\end{figure*}

To calculate the control fidelity, we approximate the unitary evolution of the time-dependent Hamiltonian $H(t;\epsilon(t))$ in (\ref{eq:system-hamiltonian}):
\begin{align*}
    U(T;\epsilon(t))
    &=\mathcal{T}\exp(-\frac{i}{\hbar}\int^{T}_{0} H(t;\epsilon(t)) dt)\\
    &\simeq \prod_{l=m}^1 \exp(-\frac{i}{\hbar}H\big((s_{l+1} + s_{l})/2,\epsilon(t)\big)\big(s_{l+1} - s_{l}\big)),
\end{align*}
where $\mathcal{T}$ is the time ordering operator. The integral is discretized into $m$ segments with boundaries $s_l$, as illustrated in Fig.~\ref{fig:fidelity-calc}(a).

To match the segment boundaries at the pulse time, we first define uniform segment boundaries $s'_l=(l-1)T/m$ for $l=1,\dots,m+1$, where $m$ is sufficiently large to ensure that the mapping from each pulse time $T_k$ to its closest boundary is one-to-one. Next, we find the set of indices of the closest boundaries $L^*=\{\underset{l}{\arg\min} |T_k-s'_l|~|~k=1,\dots,n\}$, and define the segment boundaries:
\begin{equation*}
s_l =
    \begin{cases}
        s'_l, & \text{if } l \notin L^* \\
        T_k~\text{where}~k = \underset{k}{\arg\min}|T_k - s'_l|, & \text{if } l \in L^*
    \end{cases}
\end{equation*}
This adjustment preserves accuracy in representing the evolution at the start and end of each pulse.

In the training process, we use 20 segments to calculate the fidelity and 100 segments for the figures presented in the main text to achieve higher accuracy. However, we observe that fidelities converge for $m\geq20$, with negligible changes for larger values, as shown in Fig.~\ref{fig:fidelity-calc}(b).

\subsection{Deep Reinforcement Learning}
\begin{figure*}
  \centering
  \includegraphics[width=1\textwidth]{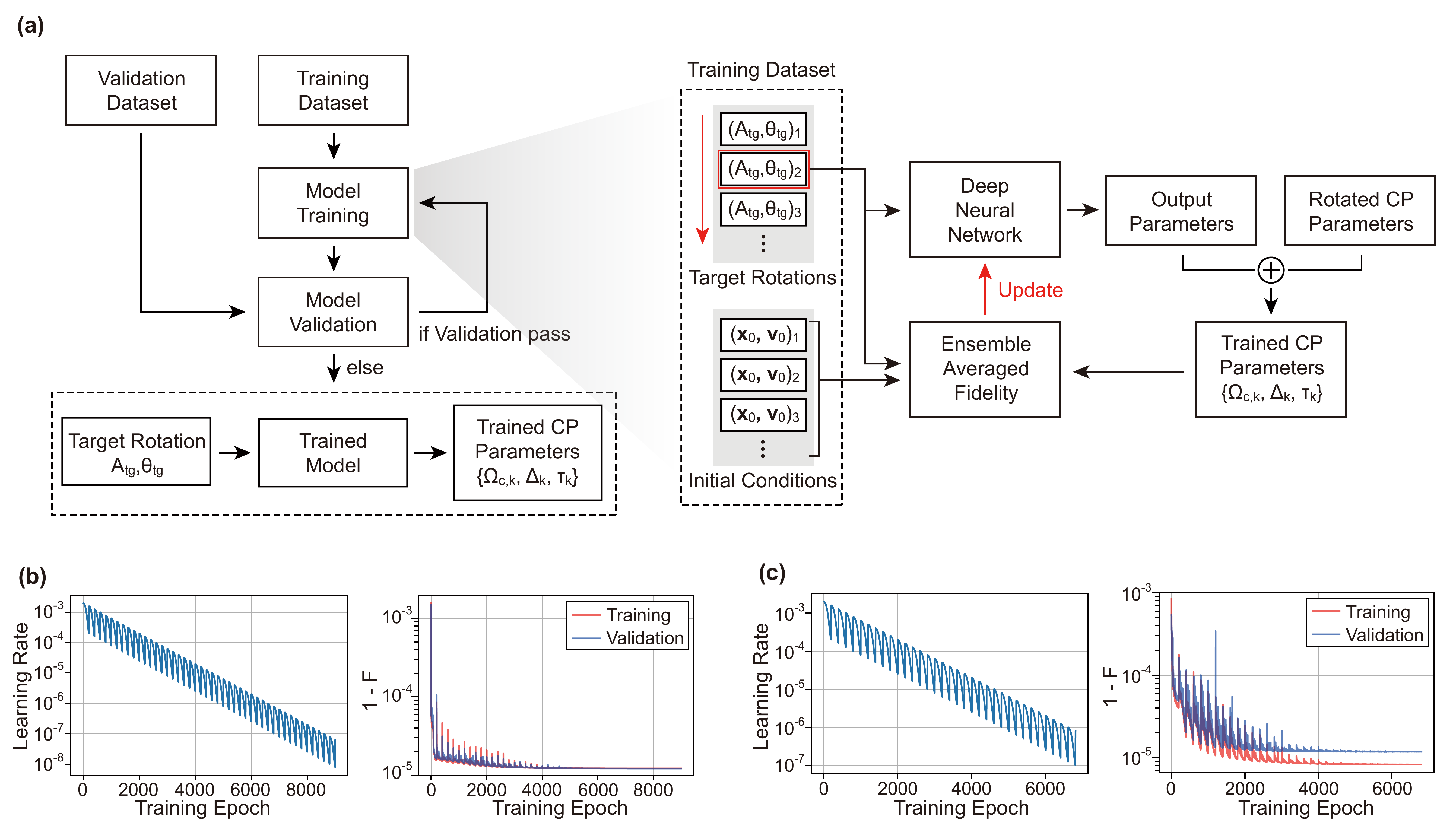}
  \caption{
    \textbf{DRL Training}
    \textbf{(a)} The flow chart of the DRL training. The left shows the total schematic of training and validation, and the right shows the update process of one epoch during the model training.
    \textbf{(b)} and \textbf{(c)} The learning rate and the fidelity during the training of \textit{CP}(3) and \textit{CP}(4), respectively. The initial learning rate is set to 0.002 and decreases by a factor of 10 every 2,000 epochs, with a cosine period of 200 epochs.
    }
  \label{fig:DRL-flow}
\end{figure*}
In this section, we describe how the deep neural network (DNN) is trained by reinforcement learning (RL) to output composite pulse (CP) parameters that suppress motion-induced errors and achieve target qubit rotations with high fidelity. Unlike multi-step RL tasks~\cite{supp_fosel_reinforcement_2018, DRL_niu_universal_2019}, our problem focuses on a single-step (open-loop) optimization~\cite{supp_viquerat_direct_2021} for each target rotation. Below, we detail the network architecture, dataset preparation, and training procedure.

Finding CP parameters $\{A_k,\dot{A}_k=dA_k/dt,\theta_k,\phi_k\}$ for a given target rotation $\{A_\text{tg}, \theta_\text{tg}\}$ is an independent, open-loop problem. We determined that a direct, gradient-based DNN update using the ensemble-averaged control fidelity as a reward is sufficient, obviating the need for more complex RL algorithms like Proximal Policy Optimization.

We employ a DNN architecture with six hidden layers, each containing 128 nodes and ELU activation functions to introduce non-linearity. It outputs a set of rotation parameters $\{A_k,\dot{A}_k=dA_k/dt,\theta_k,\phi_k\}$ which maps into the pulse sequence parameters $\{\Omega_{\text{c},k},\Delta_k,\tau_k\}$ (see Sec.~\ref{section:pulse-description} for details of mapping).

To expedite training and avoid poor local minima, we initialize the DNN outputs close to conventional composite pulses (\textit{SK}1 or \textit{BB}1). This is achieved by scaling the randomly initialized outputs down by a factor of 4 and adding them to the rotation parameters of conventional CPs. For $\theta_\text{tg} \neq \pi/2$, where conventional CPs are not defined, we extended their definition by applying an appropriate rotation to each rotation axis of CP (as detailed in Sec.~\ref{section:rotated-composite-pulses}).

Figure \ref{fig:DRL-flow}(a) illustrates the flow of the training process. Before training, we create a training dataset and a validation dataset. Each dataset consists of target rotations $\{A_{\rm tg}, \theta_{\rm tg}\}$ and initial conditions $\{\boldsymbol{x}_0,\boldsymbol{v}_0\}$ of the thermal atom. For the training dataset, we uniformly sample 48 values of $A_{\rm tg} \in [\pi/4, \pi]$ and 32 values of $\theta_{\rm tg} \in [\pi/5, 4\pi/5]$, a total of $48\times 32$ target rotations, along with 128 initial conditions. For the validation dataset, we randomly sample 128 target rotations within the same range and 64 initial conditions, ensuring that there are no overlaps with the training dataset.

During the training, we iterate over all target rotations in the training dataset (in mini-batches of size 32) and compute the ensemble-averaged fidelity for each batch. The weights and biases of the DNN are updated via backpropagation to maximize the fidelity. We use the Adam optimizer with a cosine-annealing learning-rate schedule (Figs.~\ref{fig:DRL-flow}(b) and (c)), reducing the rate every 2,000 epochs. After each epoch, the DNN is evaluated against the validation dataset to monitor for overfitting. If the validation fidelity fails to improve for 1,000 consecutive epochs, we terminate training and restore the best-performing checkpoint. The training was conducted on a local workstation equipped with a single NVIDIA RTX 4090 GPU (24 GB memory). The deep neural network was trained for 8000 epochs over approximately 24 hours using built-in functions in PyTorch 2.6 with CUDA 12.6 support. Random seeds were fixed to ensure reproducibility.

\subsection{Pulse Description\label{section:pulse-description}}
\begin{figure}
  \centering
  \includegraphics[width=0.5\textwidth]{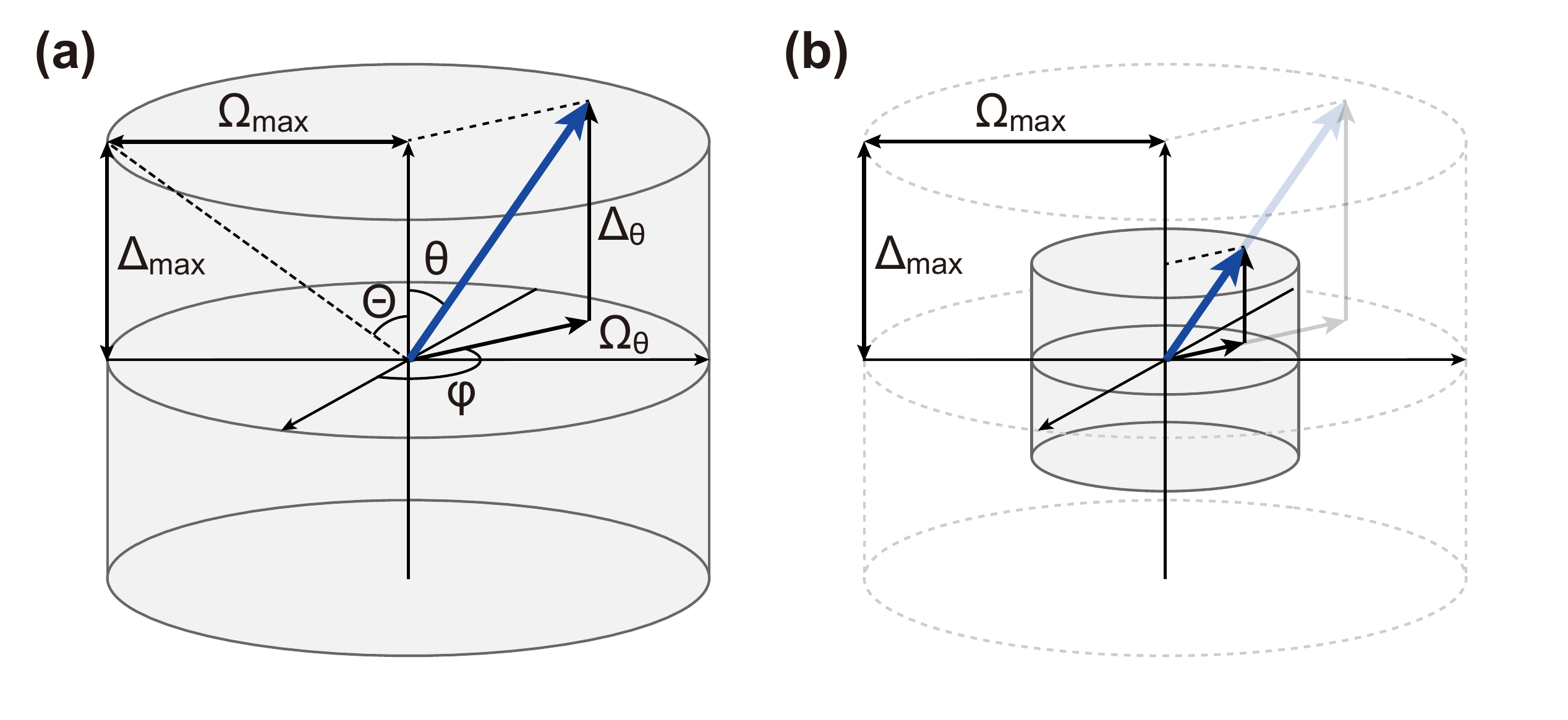}
  \caption{
    \textbf{Pulse Description}
    \textbf{(a)} The surface of the cylinder represents the maximum boundary of rotation vector in SO(3) representation. The blue arrow indicates the maximum available rotation vector for arbitrary $\theta$ and $\phi$.
    \textbf{(b)} The gray cylinder surface represents the minimum boundary of rotation vector, scaled down by a factor of 0.1 from the left figure.
    }
  \label{fig:pulse-description}
\end{figure}

Here, we describe the mathematical relationship between the rotation parameters $\{A,\dot{A},\theta,\phi\}$ which are the outputs of the DNN, and the pulse parameters $\{\Omega_\text{c}, \Delta, \tau\}$. To prevent the trained CPs from being unphysically fast, we impose the upper bound of $2\pi\times 1~\text{MHz}$ on both the Rabi frequency $\Omega_\text{max}$ and detuning $\Delta_\text{max}$, as shown in Fig.~\ref{fig:pulse-description}(a).

From the output polar angle $\theta$, the maximum available Rabi frequency $\Omega_\theta$ and detuning $\Delta_\theta$ are determined by comparing $\theta$ with a threshold angle $\Theta = \arctan(\Omega_\text{max}/\Delta_\text{max})$:

\begin{align*}
\Omega_\theta =
\begin{cases}
\Delta_\text{max} \tan\theta, & \theta \le \Theta,\\[6pt]
\Omega_\text{max}, & \Theta < \theta \le \pi - \Theta,\\[6pt]
\Delta_\text{max} \tan\theta, & \theta > \pi - \Theta,
\end{cases}
\qquad
\Delta_\theta =
\begin{cases}
\Delta_\text{max}, & \theta \le \Theta,\\[6pt]
\Omega_\text{max} / \tan\theta, & \Theta < \theta \le \pi - \Theta,\\[6pt]
-\Delta_\text{max}, & \theta > \pi - \Theta.
\end{cases}
\end{align*}

The pulse parameters are defined as
\begin{align*}
    \Omega_\text{c} = \chi \Omega_\theta e^{i\phi},
    \quad
    \Delta = \chi \Delta_\theta,
    \quad
    \tau = A / \dot{A},
\end{align*}
where $\chi = \dot{A} / \sqrt{\Omega_\theta^2 + \Delta_\theta^2}$ is the scaling factor between the desired control speed $\dot{A}$ and its maximum value $\sqrt{\Omega_\theta^2 + \Delta_\theta^2}$. To ensure experimentally feasible operation, we constrained $\dot{A}$ such that $\chi \in [0.1,1]$. Therefore, the rotation vector resides between the cylinder surfaces in Figs.~\ref{fig:pulse-description}(a) and (b).

\subsection{Rotated Composite Pulses\label{section:rotated-composite-pulses}}
\begin{figure*}
  \centering
  \includegraphics[width=1\textwidth]{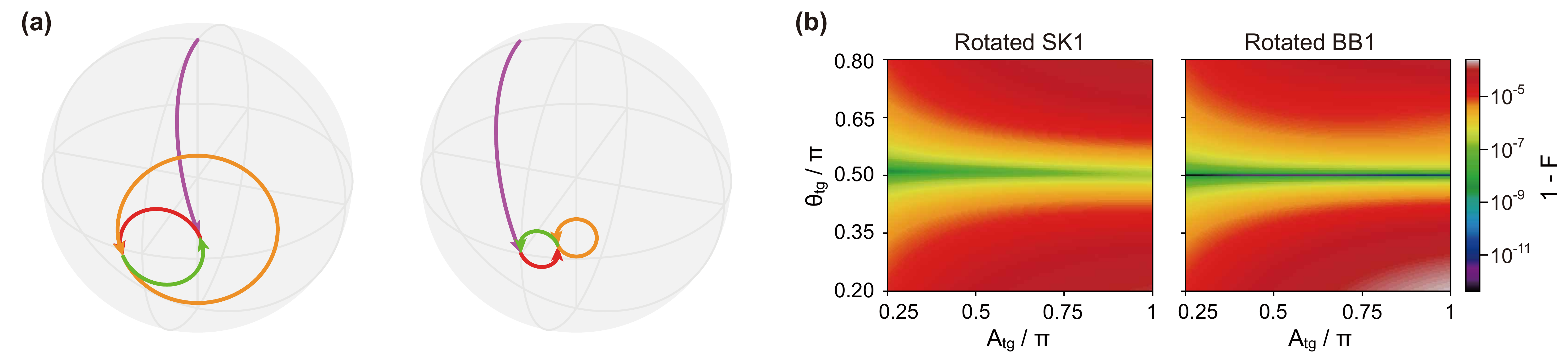}
  \caption{
    \textbf{Rotated Composite Pulses}
    \textbf{(a)} The state trajectory produced by the newly defined rotated \textit{BB}1 for a $\theta_{\rm tg} \neq \pi/2$ (left). Since the rotated \textit{BB}1 is derived by rotating the rotation axes of original \textit{BB}1, its trajectory is equivalent to that of the original \textit{BB}1 if the same rotation is applied to the initial state (right).
    \textbf{(b)} The control fidelity of rotated \textit{SK}1 (left) and rotated \textit{BB}1 (right) in the presence of 1\% fractional static amplitude error. The robustness performance decreases rapidly as $\theta_{\rm tg}$ deviates from the $\pi/2$.
  \label{fig:rotated-CPs}
  }
\end{figure*}

To use the conventional CP as an initial DNN output, even for cases where $\theta_{\rm tg} \neq \pi/2$, we replace the rotation vector $\hat{\mathbf{n}}$ of each pulse of CP for arbitrary $\theta_{\rm tg}$:
\begin{align*}
    \hat{\mathbf{n}}(\theta_k, \phi_k) \rightarrow R_y(\theta_{\rm tg} - \pi/2)\hat{\mathbf{n}}(\theta_k, \phi_k)
\end{align*}
where $R_y(\theta)$ is a SO(3) rotation matrix along the y axis. Figure~\ref{fig:rotated-CPs}(a) shows the trajectory of the rotated \textit{BB}1 on the Bloch sphere. As shown in Fig.~\ref{fig:rotated-CPs}(b), the rotated conventional CPs are no longer robust when $\theta_{\rm tg}$ deviates from $\pi/2$.

\subsection{Error Analysis}
\begin{table}
  \centering
  \setlength{\tabcolsep}{10pt} % Default value: 6pt
  \renewcommand{\arraystretch}{1.5} % Default value: 1
  \begin{tabular}{ l l c }
  \hline \hline
    Source & Type & Control Error \\
    \hline \hline
    \multirow{1}{*}{Motion Induced Error}
    &Amplitude & $\mathcal{O}(10^{-3})$ \\
    \hline
    \multirow{1}{*}{Differential Light Shift}
    &Detuning & $\mathcal{O}(10^{-6})$ \\
    \hline
    \multirow{1}{*}{Polarization Mixing}
    &Loss & $\mathcal{O}(10^{-5})$ \\
    \hline
    \multirow{1}{*}{Incoherent Scattering}
    &Loss & $\mathcal{O}(10^{-4})$ \\
    \hline \hline
  \end{tabular}
  \caption{
  \textbf{Error Budget} A summary of dominant error sources for a $\pi$-pulse implemented via two-photon Raman transitions.
  }
  \label{tab:error-budget}
\end{table}

\begin{table}
  \centering
  \setlength{\tabcolsep}{12pt}
  \renewcommand{\arraystretch}{1.5}
  \begin{tabular}{lr}
    \hline \hline
    Tweezer Beam Radius 
    ($1/e^2$) & $0.7~\upmu\mathrm{m}$ \\
    Trap Depth & $0.8~{\rm mK}$ \\
    Trap Frequency $\omega_r$ $(\omega_z)$& $ 2\pi\times155$ $(42)~{\rm kHz}$ \\
    \hline
    Atom Temperature & $30~{\rm \upmu K}$ \\
    Atom Distribution $\sigma_r~(\sigma_z)$ & $78~(201)~{\rm nm}$\\
    \hline \hline
  \end{tabular}
  \caption{
  \textbf{Parameters for the Trap and Atom Motions}  $\sigma_r~(\sigma_z)$ denotes the standard deviation of the atom's position along the radial (axial) direction.
  }
  \label{tab:parameters}
\end{table}

The unitary evolution of a single rectangular pulse can be described by the pulse parameters:
\begin{align*}
    U &= I\cos{\dfrac{\tilde{\Omega}T}{2}}-\dfrac{i}{\tilde{\Omega}}
    \begin{pmatrix}
      \Delta & \Omega^* \\
      \Omega & -\Delta
    \end{pmatrix}
    \sin{\dfrac{\tilde{\Omega}T}{2}},
\end{align*}
where $\Omega$ is the Rabi frequency, $\Delta$ is the detuning, $T$ is the pulse time, and $\tilde{\Omega}=\sqrt{|\Omega|^2+\Delta^2}$.

For the rectangular pulse with pulse parameters $\{\Omega_\text{c},\Delta_\text{c},T_\text{c}\}$, the control error is given by:
\begin{align*}
    1 - F &= 1 - \dfrac{1}{4}\left|{\rm Tr}(U^\dagger_\text{c} U)\right|^2 \\
      &= \left|\cos\dfrac{\tilde{\Omega}_\text{c}T_\text{c}}{2}\cos\dfrac{\tilde{\Omega}T}{2}
      +\dfrac{\Omega_\text{c} \Omega+\Delta_\text{c}\Delta}{\tilde{\Omega}_\text{c}\tilde{\Omega}}
      \sin\dfrac{\tilde{\Omega}_\text{c}T_\text{c}}{2}\sin\dfrac{\tilde{\Omega}T}{2}\right|^2,
\end{align*}
where $U_\text{c}$ denotes the ideal time-evolution with the pulse parameters and $U$ is the actual time-evolution including errors.

In this analysis, we neglect errors in pulse length and phase of the Rabi frequency, i.e., $T=T_\text{c}$ and $\arg({\Omega}) = \arg({\Omega_\text{c}})$. We also assume a real Rabi frequency, i.e., $\Omega_\text{c}^*=\Omega_\text{c}$. We present an error budget (Table~\ref{tab:error-budget}) for resonant rectangular pulses ($\Delta_\text{c} =0$) calculated under the conditions specified in Table~\ref{tab:parameters} and assuming a control beam with a $1/e^2$ radius of $1~\upmu\text{m}$.

\medskip
\subsubsection{Motion Induced Amplitude Error}
The detailed formulas are provided in Sec.~\ref{section:motion-induced-amplitude-error}. Here, we summarize the estimation of the control error:
\begin{align*}
    1 - F &= 1 - \left|1 - \frac{1}{8}\epsilon^2\Omega_\text{c}^2T^2 + \mathcal{O}(\epsilon^4)\right|^2\\
    &\simeq \frac{1}{4}(\epsilon\Omega_\text{c}T)^2,
\end{align*}
where $\epsilon = (\Omega - \Omega_\text{c})/\Omega_\text{c}$ is the fractional amplitude error induced by atom motion. By averaging over the thermal position distribution of the atom given in (\ref{eq:initial-condition}), the expected control error is approximately $9\times 10^{-4}$ per $\pi$-pulse.

\medskip
\subsubsection{Differential Light Shift}
The qubit basis states $|F=1, m_F=0\rangle$ and $|F=2, m_F=0\rangle$ of $\rm 5^2 S_{1/2}$ manifold of $^{87}$Rb exhibit different polarizabilities for the tweezer and the control beam. As the atom oscillates within the trap, these states experience different AC Stark shifts, resulting in a differential light shift. This shift modifies the detuning experienced by the atom, thereby inducing a detuning error that impacts qubit control fidelity. In the presence of detuning error the control fidelity can be approximated by: 
\begin{align*}
    1 - F &= 1 - \left|1 - \frac{1}{2}\epsilon_\text{d}^2\sin^2\frac{\Omega_\text{c}T}{2}+ \mathcal{O}(\epsilon_\text{d}^4)\right|^2\\
    &\simeq \left(\epsilon_\text{d}\sin\frac{\Omega_\text{c}T}{2}\right)^2
\end{align*}
where $\epsilon_\text{d} \doteq \Delta_\text{LS}/\Omega_\text{c}$ is the fractional detuning error, where $\Delta_\text{LS}$ is the induced detuning due to the differential light shift.

\begin{figure}
  \centering
  \includegraphics[width=0.5\textwidth]{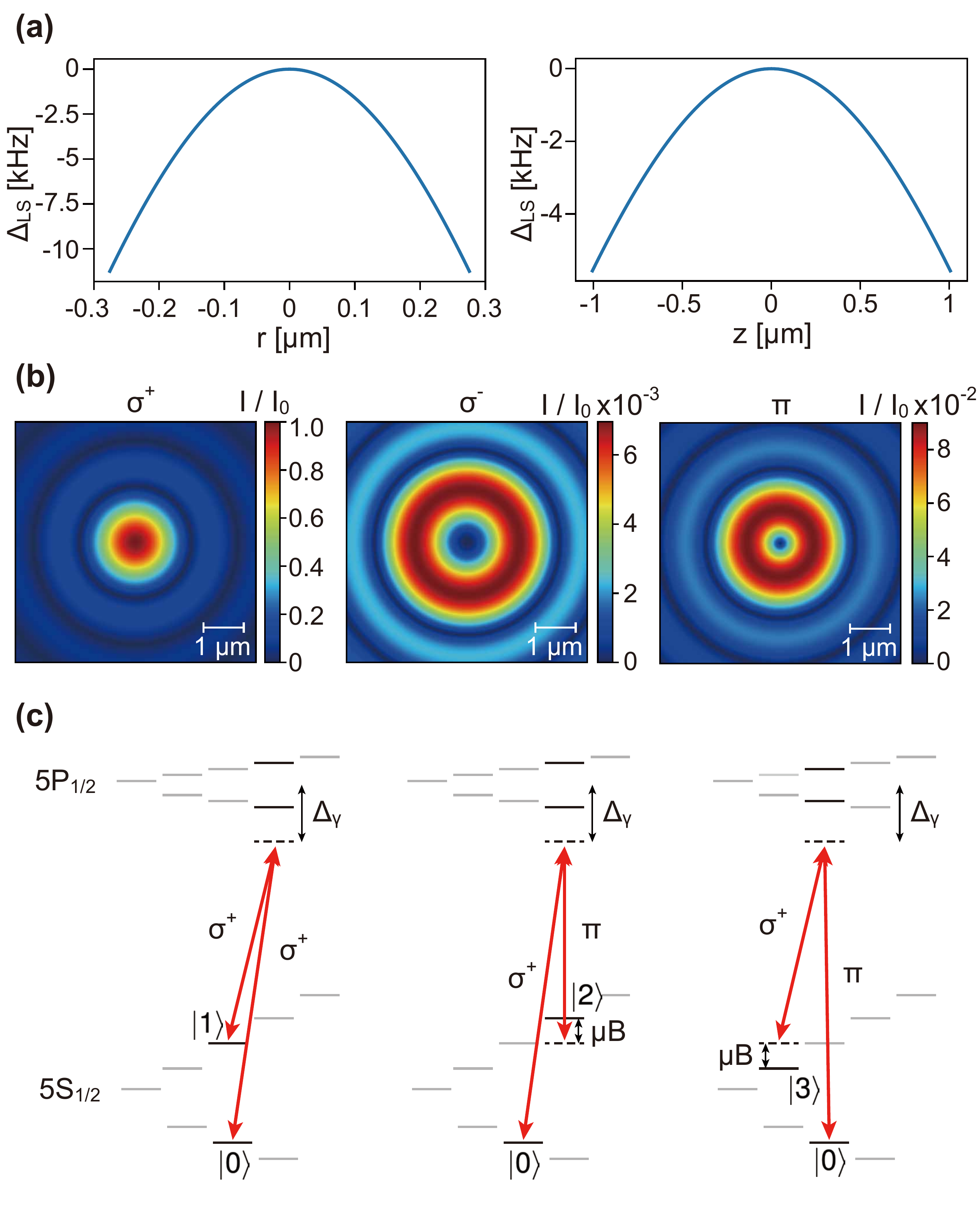}
  \caption{
    \textbf{Error Budgeting}
    \textbf{(a)} Induced detuning due to differential light shift in radial (left) and axial (right) directions.
    \textbf{(b)} Intensity in focal plane for each polarization component normalized by peak intensity of $\sigma^+$ polarization.
    \textbf{(c)} The three possible pathways of Raman transitions. The $\pi$ polarization component induces population transfer into non-qubit states $|2\rangle=|F=2, m_F=1\rangle$ and $|3\rangle=|F=2, m_F=-1\rangle$.
    }
  \label{fig:error-budget}
\end{figure}

To quantify the differential light shift, we calculate the dynamical polarizability of $|F=1, m_F=0\rangle$ and $|F=2, m_F=0\rangle$ clock states in the $\rm 5^2 S_{1/2}$ manifolds of $^{87}$Rb. Our calculation employs the theoretical expressions provided in \cite{intro_advantage_steck_quantum_2023} and atomic data from the Alkali Rydberg Calculator (ARC) Python package. Figure \ref{fig:error-budget}(a) shows the spatial distribution of $\Delta_\text{LS}$. The detuning is induced by the fields of an 852~nm tweezer and a control beam. As illustrated in the left diagram of Fig.~\ref{fig:error-budget}(c), the control beam consists of two 795~nm Raman beams. These beams have a frequency difference of $2\pi \times 6.8~{\rm GHz}$ corresponding to the ground-state hyperfine splitting, and are collectively detuned by $\Delta_\gamma=2\pi \times  100~{\rm GHz}$ from the $5\text{P}_{1/2}$ manifold. By averaging $\Delta_\text{LS}$ over the thermal position distribution of the atom, we estimate the ensemble-averaged control error for a $\pi$-pulse to be approximately $2\times 10^{-6}$.

\medskip
\subsubsection{Polarization Mixing}

When a laser beam passes through a high-numerical-aperture lens, its local polarization vector rotates along with the converging wave vectors. This results in a mixing of polarization components in the focal region. An atom oscillating within a focused control beam can interact with these unintended polarization components, causing unwanted coupling to non-qubit states.

To estimate the population loss induced by this polarization mixing, we calculate the local polarization components near the focus using vector diffraction theory, specifically the Debye-Wolf integral~\cite{DRL_budget_joseph_braat_imaging_2019}. We model a $\sigma^+$-polarized plane wave at 795~nm focused by a lens with a numerical aperture of 0.345, which produces a focal spot with a $1/e$ radius of $1~\upmu\text{m}$. Figure \ref{fig:error-budget}(b) shows the calculated intensity distribution of each polarization component in the focal plane.

In our Raman scheme driven by $\sigma^+$-$\sigma^+$ transitions, the dominant population loss arises from the off-resonant $\sigma^+$-$\pi$ and $\pi$-$\sigma^+$ pathways, as shown in Fig.~\ref{fig:error-budget}(c). To quantify this effect, each of these error channels is modeled as an independent, effective two-level system. Contributions from any residual $\sigma^-$ polarization are considered negligible, and $\pi$-$\pi$ transitions are excluded as they are forbidden by dipole selection rules. 

For each effective two-level system the Rabi frequency and detuning are given by:
\begin{align*}
    &\Omega_{|0\rangle \leftrightarrow |1\rangle} = \Omega_\text{c},&&\Delta_{|0\rangle \leftrightarrow |1\rangle} = 0\\
    &\Omega_{|0\rangle \leftrightarrow |2\rangle} = \xi\Omega_\text{c},&&\Delta_{|0\rangle \leftrightarrow |2\rangle} = (1-\xi^2)\frac{\Omega_\text{c}}{2} + \mu B\\        
    &\Omega_{|0\rangle \leftrightarrow |3\rangle} = \xi\Omega_\text{c},&&\Delta_{|0\rangle \leftrightarrow |3\rangle} = (\xi^2-1)\frac{\Omega_\text{c}}{2} - \mu B
\end{align*}
where $\xi$ is the amplitude ratio between the undesired $\pi$ polarization and the intended $\sigma^+$ polarization, $\mu = 0.7~{\rm MHz/G}$ is the magnetic moment of the $|F=2\rangle$ manifold in $\rm 5^2 S_{1/2}$, and $B$ is the external magnetic field along the beam propagation axis.

The population loss due to polarization mixing is estimated as the incoherent sum of the maximum transition probabilities to $\ket{2}$ and $\ket{3}$:
\begin{align*}
    \sum_{i= 2, 3}\left|\dfrac{\Omega_{|0\rangle \leftrightarrow |i\rangle}}{\sqrt{\Omega_{|0\rangle \leftrightarrow |i\rangle}^2 + \Delta_{|0\rangle \leftrightarrow |i\rangle}^2}}\right|^2 = 2\left|\dfrac{2\xi}{1+2\mu B / \Omega_\text{c}} + \mathcal{O}(\xi^3)\right|^2.
\end{align*}
This estimation yields the same order of magnitude as the numerical solution of the four-level system ($\ket{0}, \ket{1}, \ket{2}\text{ and } \ket{3}$) in the regime of $\xi<10^{-1}$ and $\mu B/\Omega_\text{c} >10$. Note that the loss can be suppressed by increasing the magnetic field strength relative to the Rabi frequency, thereby enhancing the detuning of the unwanted transitions. For $B=10~{\rm G}$ and $\xi$ determined from Fig.~\ref{fig:error-budget}(b), the ensemble-averaged population loss is estimated to be approximately $9\times10^{-6}$ for resonant pulses.

\medskip
\subsubsection{Incoherent Scattering}

During the Raman transitions, the residual population in $5\text{P}_{1/2}$ manifold can decay incoherently to the other ground state, a process known as incoherent scattering. The occurrence probability of this process is determined by the product of the excited state's decay rate, its population and the pulse time:
\begin{align*}
    \Gamma \max(\rho_\text{ee})T = \Gamma \frac{\Omega_\text{c}}{2\Delta_\gamma}T
\end{align*}
where $\Gamma$ is the spontaneous decay rate of $5\text{P}_{1/2}$ manifold. In the regime specified in Table \ref{tab:parameters}, with $\Delta_\gamma = 2\pi \times  100~{\rm GHz}$, the population loss due to incoherent scattering is estimated to be approximately $9\times10^{-5}$ per $\pi$-pulse.

\begin{figure*}[p]
  \centering
  \includegraphics[width=1\textwidth]{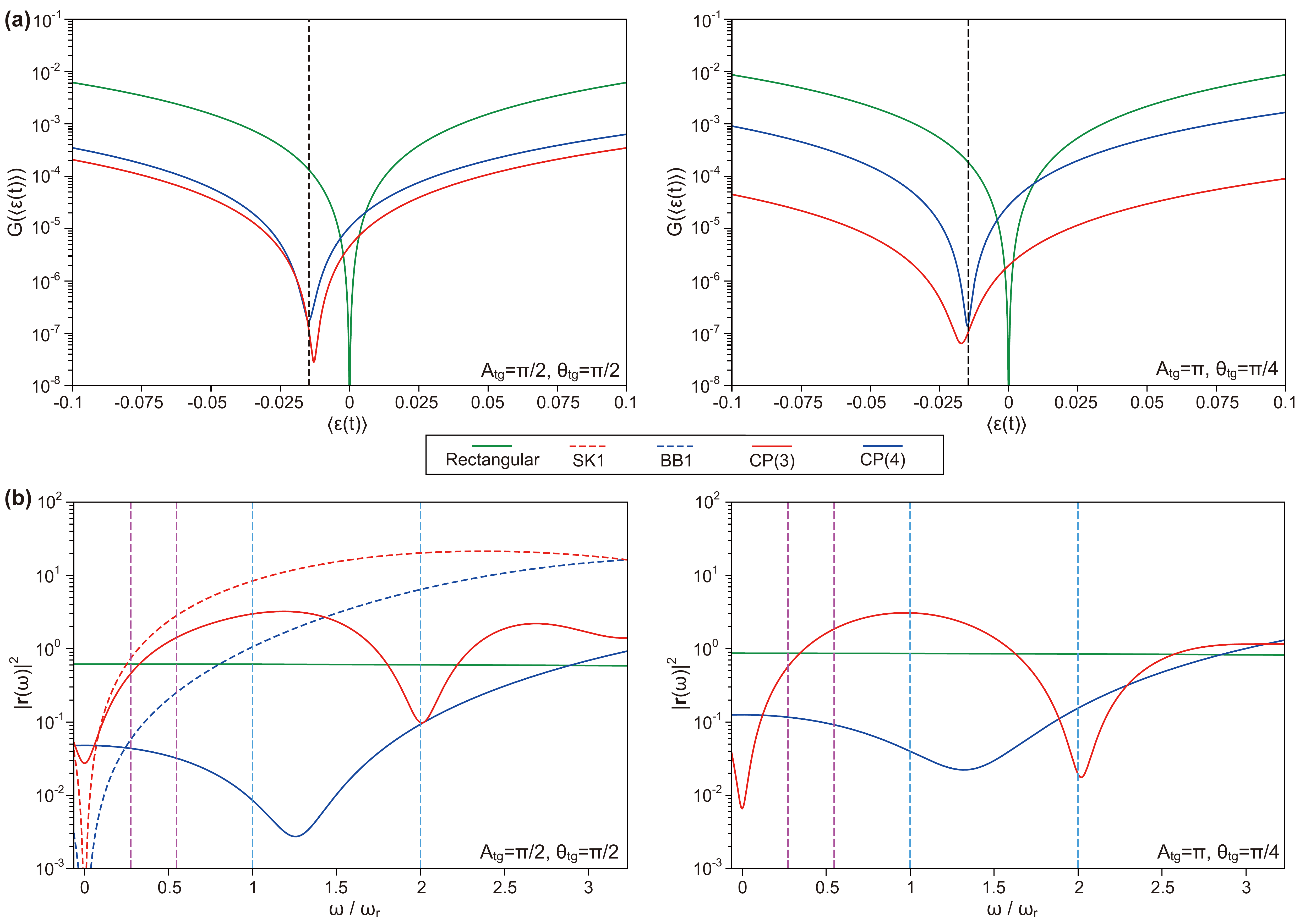}
  \caption{
    \textbf{Spectral Characteristics for Other Target Rotations}
    \textbf{(a)} The residual bias of the CPs for a $\pi/2$-pulse (left) and a Hadamard gate (right). The black dotted line indicates the error bias for an ideal focus.
    \textbf{(b)} Filter functions of the CPs encoding a $\pi/2$-pulse (left) and a Hadamard gate (right). The trained CPs exhibit consistent characteristics across different target rotations: CP(3) provides narrowband suppression, while CP(4) provides broadband suppression. The cyan (magenta) dotted lines indicate $\omega=\omega_{r}$ and $2\omega_{r}$ ($\omega_{z}$ and $2\omega_{z}$).
    }
  \label{fig:spectral}
\end{figure*}
\clearpage

\bibliography{main}